\documentclass[11pt]{article}
\usepackage{graphicx} 
\usepackage{float} 
\usepackage{amssymb}
\usepackage{amsmath}
\usepackage{mathtools}
\usepackage[thmmarks,amsmath]{ntheorem} 
\usepackage{bm} 
\usepackage[colorlinks,linkcolor=cyan,citecolor=cyan]{hyperref}
\usepackage{extarrows}
\usepackage[hang,flushmargin]{footmisc} 
\usepackage{mathrsfs} 
\usepackage[font=footnotesize,skip=0pt,textfont=rm,labelfont=rm]{caption,subcaption}
\usepackage{booktabs} 
\usepackage{tocloft}
\usepackage[]{longtable}
\usepackage{array}
\usepackage{comment}
{
    \theoremstyle{nonumberplain}
    \theoremheaderfont{\bfseries}
    \theorembodyfont{\normalfont}
    \theoremsymbol{\mbox{$\Box$}}
    
}

\usepackage{theorem}
\newtheorem{theorem}{Theorem}
\newtheorem{lemma}{Lemma}
\newtheorem{definition}{Definition}
\newtheorem{assumption}{Assumption}

{
    \theoremheaderfont{\bfseries}
    \theorembodyfont{\normalfont}
    
}
\numberwithin{equation}{section} 

\usepackage{chngcntr}
\usepackage[shortlabels]{enumitem}
\usepackage{apptools}
\AtAppendix{\counterwithin{lemma}{section}}
\AtAppendix{\counterwithin{theorem}{section}}
\AtAppendix{\counterwithin{corollary}{section}}
\AtAppendix{\counterwithin{assumption}{section}}
\AtAppendix{\counterwithin{definition}{section}}
\AtAppendix{\counterwithin{table}{section}}
\AtAppendix{\counterwithin{footnote}{section}}
\usepackage[a4paper]{geometry}                           
\usepackage{adjustbox}
\usepackage[figuresright]{rotating}
\usepackage{float}
\usepackage{multirow}

\newcommand{\cov}{\text{cov}}
\newcommand{\indep}{\perp \!\!\! \perp}

\usepackage{natbib}
\usepackage{changepage}

\date{First version: March 11, 2021 \\ This version: \today}
\begin{document}
\title{\textbf{More Robust Estimators for Instrumental-Variable Panel Designs, With An Application to the Effect of Imports from China on US Employment.}\thanks{We are particularly grateful to Xavier D'Haultf\oe{}uille, Peter Hull, Michal Koles\'ar, and Isabelle Méjean for their feedback on this paper. We also thank Teresa Fort, Lucie Gadenne, François Gerard, Paul Goldsmith-Pinkham, Shawn Klimek, Ismael Mourifi\'e, F\'elix Pasquier, Aureo de Paula, Jonathan Roth, Isaac Sorkin, Martha Stinson, and seminar participants at CREST, McMaster University, PSE, Queen Mary, Tilburg University, the Tinbergen Institute, and the Sao Paulo School of Economics for their helpful comments. Cl\'{e}ment de Chaisemartin was funded by the European Union (ERC, REALLYCREDIBLE,GA N°101043899). Views and opinions expressed are those of the authors and do not reflect those of the European Union or the European Research Council Executive Agency. Neither the European Union nor the granting authority can be held
responsible for them.}}

\author{Cl\'ement de Chaisemartin\footnote{Economics Department, Sciences Po, clement.dechaisemartin@sciencespo.fr.} \and Ziteng Lei\footnote{School of Labor and Human Resources, Renmin University of China, leiziteng@ruc.edu.cn.}}
\maketitle
\thispagestyle{empty}

\begin{abstract}
We show that first-difference two-stages-least-squares regressions identify non-convex combinations of location-and-period-specific treatment effects. Thus, those regressions could be biased if effects are heterogeneous. We propose an alternative instrumental-variable correlated-random-coefficient (IV-CRC) estimator, that is more robust to heterogeneous effects. We revisit Autor et al. (2013), who use a first-difference two-stages-least-squares regression to estimate the effect of imports from China on US manufacturing employment. Their regression estimates a highly non-convex combination of effects. Our more robust IV-CRC estimator is small and insignificant. Though its confidence interval is wide, it significantly differs from the first-difference two-stages-least-squares estimator.
\end{abstract}



\newpage
\section{Introduction}

First-difference two-stage-least-squares (FD 2SLS) regressions are a popular tool to estimate the effect of a treatment on an outcome. For instance, \citet{autor2013china}, herafter ADH, use a panel data set of US commuting zones (CZs) to estimate the effect of $D_{g,t}$, the imports from China in CZ $g$ at $t$,\footnote{ADH's treatment is actually a proxy for $g$'s imports from China at $t$. The simplified description of their treatment we give in this introduction is not of essence to our main conclusions.} on $Y_{g,t}$, the manufacturing employment in $g$ at $t$. Some of their regressions leverage two time periods per CZ, while others leverage three periods: to simplify the exposition without great loss of generality, we assume the data has two periods in this introduction. Then, one may estimate an OLS regression of $\Delta Y_g$ on $\Delta D_g$, where $\Delta$ denotes the FD operator. However, $\Delta D_g$ may be endogenous: the evolution of imports from China may be correlated with US demand shocks. Therefore, ADH use an instrument $Z_{g,t}$, whose construction we detail below, and run a 2SLS regression of $\Delta Y_{g}$ on $\Delta D_{g}$ using $\Delta Z_{g}$ as the instrument.  One can show that $\hat{\theta}^{b}$, the coefficient of $\Delta D_{g}$, has the following expression:
	\begin{align}
	\hat{\theta}^{b}
	&=\frac{\sum_{g=1}^G \Delta Y_{g} \left( \Delta Z_{g}-{\Delta Z}_{.}\right) }{\sum_{g=1}^G \Delta D_{g} \left( \Delta Z_{g}-{\Delta Z}_{.}\right) }\label{eq:estimator},
	\end{align}
where ${\Delta Z}_{.}$ is the average of $\Delta Z_{g}$ across CZs. With two periods, $\hat{\theta}^{b}$ is numerically equivalent to the coefficient of $D_{g,t}$ in a 2SLS two-way fixed effects regression (TWFE) of $Y_{g,t}$ on $D_{g,t}$ with location and period fixed effects, using $Z_{g,t}$ as the instrument. $\hat{\theta}^{b}$ has been used by several other influential papers, see e.g. \cite{autor2020importing} or \cite{acemoglu2020robots}.

\medskip
We start by showing that $\hat{\theta}^{b}$ does not estimate a convex combination of location-and-period-specific treatment effects. Our two first results are simple enough to state finite-sample versions of them in this introduction. Let $Y_{g,t}(d)$ denote the potential outcome of location $g$ at period $t$ if $D_{g,t}$ is equal to $d$. For instance, in ADH $Y_{g,t}(0)$ is CZ $g$'s potential manufacturing employment at $t$ without any imports from China. We assume that
		\begin{align*}
			Y_{g,t}(d)=Y_{g,t}(0)+\alpha_{g,t} d,
		\end{align*}
meaning that $g$'s potential outcome at $t$ is a linear function of its treatment level, with a location-and-period-specific slope $\alpha_{g,t}$. Then, the observed outcome satisfies
		\begin{align*}
			Y_{g,t}=Y_{g,t}(0)+\alpha_{g,t} D_{g,t}.
		\end{align*}
First-differencing the previous display yields
\begin{align}\label{eq:obs_outcome}
		\Delta Y_{g} = \Delta Y_{g} (0) + \alpha_{g,2} D_{g,2}-\alpha_{g,1} D_{g,1}=\Delta Y_{g} (0) + \alpha_{g,2} \Delta D_{g}+\Delta \alpha_g D_{g,1}.
\end{align}
If the treatment effect is constant over time (i.e. $\alpha_{g,2}=\alpha_{g,1}=\alpha_{g}$), \eqref{eq:obs_outcome} simplifies to
\begin{align}\label{eq:obs_outcome2}
		\Delta Y_{g} = \Delta Y_{g} (0) + \alpha_{g} \Delta D_{g}.
		\end{align}
Now, plugging \eqref{eq:obs_outcome} into \eqref{eq:estimator} yields
\begin{align}
	\hat{\theta}^{b}
	&=\frac{\sum_{g=1}^G \Delta Y_{g} (0) \left( \Delta Z_{g}-{\Delta Z}_{.}\right) }{\sum_{g=1}^G \Delta D_{g} \left( \Delta Z_{g}-{\Delta Z}_{.}\right) }+\frac{\sum_{g=1}^G \left(\alpha_{g,2} D_{g,2}-\alpha_{g,1} D_{g,1}\right) \left( \Delta Z_{g}-{\Delta Z}_{.}\right) }{\sum_{g=1}^G \Delta D_{g} \left( \Delta Z_{g}-{\Delta Z}_{.}\right) }\nonumber\\
	&=\frac{\sum_{g=1}^G \Delta Y_{g} (0) \left( \Delta Z_{g}-{\Delta Z}_{.}\right) }{\sum_{g=1}^G \Delta D_{g} \left( \Delta Z_{g}-{\Delta Z}_{.}\right) }\nonumber\\
&+\sum_{g=1}^G \sum_{t=1}^2 \frac{(1\{t=2\}-1\{t=1\})D_{g,t}(\Delta Z_{g}-{\Delta Z}_{.})}{ \sum_{g'=1}^G \sum_{t'=1}^2 (1\{t'=2\}-1\{t'=1\})D_{g',t'}(\Delta Z_{g'}-{\Delta Z}_{.})}\alpha_{g,t}.\label{eq:decompo1}
	\end{align}
Thus, $\hat{\theta}^{b}$ can be decomposed into the sum of two terms. The first is the coefficient one would get from a 2SLS regression of $\Delta Y_{g} (0)$, locations' outcome evolution without treatment, on $\Delta D_{g}$, using $\Delta Z_{g}$ as the instrument. If locations' outcome evolutions without treatment are uncorrelated with $\Delta Z_{g}$, a kind of parallel-trends assumption, this term converges to zero. The second term is a weighted sum of the location-and-period-specific slopes $\alpha_{g,t}$, where weights sum to one, but where every location is such that either its period-one or its period-two slope is weighted negatively: $\alpha_{g,1}$ is weighted negatively if $\Delta Z_{g}>{\Delta Z}_{.}$, and $\alpha_{g,2}$ is weighted negatively if $\Delta Z_{g}<{\Delta Z}_{.}$. Negative weights may be problematic. Because of them, one could have, say, $\alpha_{g,t}\geq 0$ for all $(g,t)$ but $\hat{\theta}^{b}<0$, even asymptotically, and even when the instrument is exogeneous. Assuming constant treatment effects over time ($\alpha_{g,2}=\alpha_{g,1}=\alpha_{g}$), \eqref{eq:decompo1} simplifies to
\begin{align}
	\hat{\theta}^{b}
	&=\frac{\sum_{g=1}^G \Delta Y_{g} (0) \left( \Delta Z_{g}-{\Delta Z}_{.}\right) }{\sum_{g=1}^G \Delta D_{g} \left( \Delta Z_{g}-{\Delta Z}_{.}\right) }+\sum_{g=1}^G \frac{\Delta D_{g}(\Delta Z_{g}-{\Delta Z}_{.})}{ \sum_{g'=1}^G \Delta D_{g'}(\Delta Z_{g'}-{\Delta Z}_{.})}\alpha_{g}.\label{eq:decompo2}
	\end{align}
Even if the instrument is exogenous, $\hat{\theta}^{b}$ still does not estimate a convex combination of effects: $\alpha_{g}$ is weighted negatively for locations such that $\Delta D_{g}$ and $\Delta Z_{g}-{\Delta Z}_{.}$ are of a different sign.

\medskip
The intuition for \eqref{eq:decompo1} and \eqref{eq:decompo2} goes as follows. \eqref{eq:estimator} shows that locations such that $\Delta Z_{g}-{\Delta Z}_{.}>0$ are used as ``treatment-group'' locations by $\hat{\theta}^{b}$: their outcome and treatment evolutions are weighted positively. On the other hand, locations such that $\Delta Z_{g}-{\Delta Z}_{.}<0$ are used as ``control-group'' locations: their outcome and treatment evolutions are weighted negatively. $\alpha_{g,1}$ enters with a negative sign in the $\Delta Y_{g}$ of treatment-group locations (see \eqref{eq:obs_outcome}), so it gets weighted negatively by $\hat{\theta}^{b}$. Similarly, $\alpha_{g,2}$ enters with a positive sign in the $\Delta Y_{g}$ of control-group locations (see \eqref{eq:obs_outcome}), so it gets weighted negatively by $\hat{\theta}^{b}$. Assuming constant effects over time, now locations' outcome evolutions are only affected by their treatment evolutions, not by their treatment levels. But if there are treatment-group locations that experienced a negative treatment evolution, the effect of this evolution enters with a negative sign in their $\Delta Y_{g}$ (see \eqref{eq:obs_outcome2}), and it gets weighted negatively by $\hat{\theta}^{b}$. Similarly, if there are control-group locations that experienced a positive treatment evolution, the effect of this evolution enters with a positive sign in their $\Delta Y_{g}$ (see \eqref{eq:obs_outcome2}), and it gets weighted negatively.

\medskip
\eqref{eq:decompo1} and \eqref{eq:decompo2} apply to any FD 2SLS regression. An important special case of FD 2SLS regressions are FD 2SLS Bartik regressions, where the instrument has a specific shift-share structure. To introduce Bartik instruments, let us again use the ADH example. Manufacturing is divided into $S$ sectors indexed by $s$. Let $Z_{s,t}$ denote imports from China in sector $s$ at $t$, in a group of high-income countries similar to the US. The instrument in ADH is
	\begin{equation*}
	Z_{g,t}=\sum_{s=1}^S Q_{s,g} Z_{s,t},
\end{equation*}
where $Q_{s,g}$ is the share sector $s$ accounts for in CZ $g$'s manufacturing employment.
$Z_{g,t}$ is correlated to $D_{g,t}$, without being directly determined by US demand.  We derive two further decomposition results, specific to FD 2SLS Bartik regressions. First, if we further assume a linear first-stage model tailored to the structure of the Bartik instrument, $\hat{\theta}^{b}$ may still not estimate a convex combination of effects, even if the treatment effect is constant over time ($\alpha_{g,2}=\alpha_{g,1}=\alpha_{g}$) and the first-stage effect of the instrument on the treatment is fully homogeneous, across sectors, locations, and time periods. Second, even if we further assume that the shocks $Z_{s,t}$ are as-good-as randomly assigned, we show that $\hat{\theta}^{b}$ may still not estimate a convex combination of effects if treatment effects vary over time.
At the same time, we also show that with randomly-assigned shocks, there is a simple fix to the negative weights problem: a slightly modified FD 2SLS Bartik estimator, where shocks are standardized by their period-specific standard deviation when constructing the instrument, estimates a convex combination of effects, even if treatment effects vary over time and across locations. In view of this simple fix, it is important to test whether shocks are as good as randomly assigned in Bartik designs. We therefore propose two novel tests of this assumption.\footnote{Pre-testing if shocks are randomly assigned could lead to a bias if the pre-test lacks power, a concern analogous to that highlighted by \cite{roth2022pretest} in difference-in-differences studies. The benefit of pre-testing may outweight the cost. In ADH, our tests are very strongly rejected, so pre-tests do not always lack power in Bartik designs.} Our tests may be more powerful than that previously proposed by \cite{borusyak2020quasi}: when we revisit ADH, our tests are rejected while theirs is not.

\medskip
We then propose an alternative to FD 2SLS regressions, the instrumental-variable correlated-random-coefficient (IV-CRC) estimator, which is inspired from \cite{chamberlain1992efficiency}. It can be used irrespective of whether the instrument has a Bartik structure or not, provided there are at least three time periods in the data. It does not require any source of random variation, and instead relies on a parallel-trends assumption. It is much more robust to heterogeneous effects than $\hat{\theta}^{b}$: it estimates the average treatment effect, a very natural target parameter, even if the treatment effect varies across locations and over time. It still imposes some restrictions on treatment effects, as it requires that they follow the same evolution over time in every location. Moreover, it relies on a stronger parallel trends assumption than $\hat{\theta}^{b}$, and it also relies on the assumption that locations' treatment effects are mean-independent of their treatments conditional on their instruments. We propose suggestive tests of those assumptions.

\medskip
Equipped with those econometrics results, we revisit the main 2SLS FD Bartik regression in ADH. Therein, the authors estimate the effect of imports from China on US manufacturing employment, and find a large negative effect. 
We start by testing the randomly-assigned shocks assumption, and find that it is strongly rejected. Under this assumption, sectoral shocks should be uncorrelated with sectors' characteristics, and in particular with sectors' average share across locations. In practice, shocks are strongly correlated with sectoral shares, even conditional on other sectors' characteristics: this is evidence that shocks are not as-good-as randomly assigned, even conditionally. Then, we decompose the regression we revisit. Our first decomposition, following \eqref{eq:decompo1}, indicates that it estimates a highly non-convex combination of CZ-and-period specific effects $\alpha_{g,t}$: nearly 50\% of effects are weighted negatively, and negative weights sum to $-0.734$. Weights are correlated with the year variable. Weights are also correlated with several CZ characteristics, and in particular with CZs' percentage employment in routine occupations, a variable likely to be correlated with CZs' treatment effects. Then, the regression could be biased if the effects $\alpha_{g,t}$ change over time and/or are correlated with characteristics weights correlate with. Our second decomposition, following \eqref{eq:decompo2}, shows that even if one assumes constant effects over time, the regression still estimates a highly non-convex combination of effects, where negative weights sum to $-0.314$. Finally, our IV-CRC estimator is small, insignificant, significantly different from the 2SLS FD Bartik estimator, and its confidence interval does not include the Bartik estimator. Given its large standard error, our estimator is compatible with a large range of effects. To sum up, we document the three following facts: i) the random-shocks assumption is rejected in this application, ii) without this assumption, the FD 2SLS Bartik estimator therein estimates a highly non-convex combination of effects with weights correlated to plausible proxies of treatment effects, and iii) our more robust IV-CRC estimator is small and insignificant. In view of these three facts and the currently available econometrics results on FD 2SLS Bartik regressions, we believe it is reasonable to draw the following conclusion: without assuming that the effect of imports from China is constant over time and across CZs, one cannot conclude, from the particular data set used by ADH, that those imports negatively affected US manufacturing employment.

\medskip
Importantly, ADH spurred a substantial body of further research. Some papers also find a negative effect of imports from China on US labor markets \citep[see e.g. ][]{autor2014trade,acemoglu2016import}, while other papers find heterogeneous effects across firms, sectors, and locations \citep[see e.g. ][]{bloom2019impact}. Our findings do not apply to those other papers: many of them do not use FD 2SLS regressions, and all of them use different data than ADH.

\medskip
The paper is organized as follows. Section 2 presents our setup. Section 3 presents our decompositions of FD 2SLS regressions. Section 4 presents our alternative IV-CRC estimator. Section 5 presents our re-analysis of ADH. Section 6 presents recommendations for practitioners. All proofs are in the appendix.

\subsection*{Related literature}\label{sec_literature}

Our paper is related to \cite{de2020two}, who derive decompositions of OLS TWFE and FD regressions under a parallel trends assumption. Our first decomposition of $\hat{\theta}^{b}$ in Theorem \ref{th_2sls_c} below is related to their Theorem 1: replacing the instrument by the treatment in our Theorem \ref{th_2sls_c} yields the same weights as in that result with two time periods. Thus, our Theorem \ref{th_2sls_c} is an extension of that result to 2SLS regressions. \cite{de2020two} had not specifically derived a decomposition of OLS TWFE regressions in the special case with two time periods. Our Theorem \ref{th_2sls_c} can be used to that effect. The closed-form expression of the weights in that special case might be of independent interest. For instance, it shows that with two periods and $D_{g,t}>0$ for all $(g,t)$, exactly 50\% of the weights attached to OLS TWFE regressions are negative, a fact not noted in \cite{de2020two}.

\medskip
Our paper is also related to \cite{de2013note} and \cite{hudson2017interpreting}, who show that difference-in-differences (DID) 2SLS regressions, a special case of the FD 2SLS regressions we consider, can identify a LATE under parallel trends assumptions on the outcome and treatment and a monotonicity condition. However, this result does not generalize beyond the special case with two groups, two periods, binary instrument, and binary treatment they consider.

\medskip
Our paper also builds upon \citet{goldsmith2020bartik}, \citet{borusyak2020quasi}, and \citet{adao2019shift}, who have studied Bartik regressions. \citet{goldsmith2020bartik} and \citet{borusyak2020quasi} have proposed two distinct ways of rationalizing instrument-exogeneity in Bartik designs, the so-called shares and shocks approaches, respectively. Following \citet{borusyak2020quasi}, \citet{adao2019shift} have shown that in the shocks approach, conventional standard errors may be misleading, and have proposed alternative standard errors. When it considers FD 2SLS Bartik regressions, our paper is not concerned with rationalizing the Bartik instrument exogeneity: our first two decompositions of $\hat{\theta}^{b}$ hold even if the instrument is not exogenous, as \eqref{eq:decompo1} and \eqref{eq:decompo2} show. Instead, our paper is concerned with the robustness of those regressions to heterogeneous effects. Heterogeneous effects is a less central issue in those papers, though \citet{goldsmith2020bartik} discuss it in an extension, \citet{borusyak2020quasi} in their online appendix, and \citet{adao2019shift} in the main sections of their paper. More recently, \cite{borusyak2023} have written a comment on our negative result for FD 2SLS Bartik regressions with randomly-assigned shocks. To preserve space, we defer a detailed discussion of the connections between our and those four papers to Section \ref{sec_Bartik}. Finally, our IV-CRC estimator is inspired from \cite{chamberlain1992efficiency}. 

\section{Setup, notation, and main definitions}\label{sec_setup}

\paragraph{Location-level panel data.} We consider a panel with $G$ locations, indexed by $g\in \{1,...,G\}$, and $T$ periods indexed by $t\in \{1,...,T\}$. We want to use this data set to estimate the effect of a treatment $D_{g,t}$ on an outcome $Y_{g,t}$.

\subsection{Potential outcomes}\label{sub_PO}

\paragraph{Potential outcomes.} Let $Y_{g,t} (d)$  denote the potential outcome that location $g$ experiences at period $t$ if $D_{g,t}=d$. $Y_{g,t}(0)$ is $g$'s outcome at $t$ without any treatment. We make a linear treatment effect assumption
	\begin{assumption}\label{as_linear_ss}
		Linear Treatment Effect: for all $(g,t) \in \{1,...,G\}\times \{1,...,T\}$, there exists $\alpha_{g,t}$ such that for any $d$:
		\begin{align}\label{eq:causal_model_levels}
			Y_{g,t}(d)=Y_{g,t}(0)+\alpha_{g,t} d.
		\end{align}
	\end{assumption}

\paragraph{Causal model in levels or in first-difference?} An implicit assumption in the potential outcome notation above is that the levels of the treatment affect the level of the outcome. One may prefer to posit a causal model in first-difference. Let $\Delta Y_{g,t} (\delta)$  denote the potential outcome evolution that location $g$ will experience from $t-1$ to $t$ if $\Delta D_{g,t}=\delta $. $\Delta Y_{g,t} (\tilde{0})$ is $g$'s potential outcome evolution without any treatment change, where we use $\tilde{0}$ instead of $0$ to emphasize that $\Delta Y_{g,t} (\tilde{0})$ is a counterfactual outcome evolution without any treatment change, rather than without any treatment. Then, one may assume that:
		\begin{align}\label{eq:causal_model_fd}
			\Delta Y_{g,t}(\delta)=\Delta Y_{g,t} (\tilde{0})+\alpha^{fd}_{g,t} \delta.
		\end{align}
Our choice of positing a causal model in levels rather than in first-difference
has consequences for our third decomposition of $\hat{\theta}^b$ in Theorem \ref{th_2sls_c_randomshocks}. There, we show that under the linear treatment-effect model in levels in \eqref{eq:causal_model_levels}, FD 2SLS Bartik regressions with randomly-assigned shocks are not robust to time-varying effects. \cite{borusyak2020quasi} and \cite{borusyak2023} instead show that under the linear treatment-effect model in first-difference in \eqref{eq:causal_model_fd}, FD 2SLS Bartik regressions with randomly-assigned shocks are robust to time-varying effects. \cite{borusyak2023} argue that economic theory often rationalizes causal models in first-difference, as in
\eqref{eq:causal_model_fd}. While there are instances where this is true, this is not the case in ADH: in Section II.A of their Web Appendix, ADH motivate their empirical specification with a small open economy model relating ``total employment in traded goods'' in a CZ (see page 7) to their treatment variable, rather than the evolution of total employment to the evolution of their treatment variable. But more importantly, even when economic theory rationalizes a model in first-difference, this is not enough to ensure that this model can have time-varying treatment effects. Having a causal model in first-difference with time-varying effects requires ruling out the possibility that there exist a causal model in levels, a very strong requirement in our opinion. For instance, in ADH, even without resorting to a model, it makes intuitive sense that different levels of imports from China in CZ $g$ at $t$ would lead to different manufacturing employment levels there. To show that having a causal model in first-difference with time-varying treatment effects requires ruling out a causal model in levels, we prove in Lemma \ref{lem:causal_fd_levels} that if one jointly imposes a causal model in levels and in first-difference, the treatment effect has to be constant over time in both models.
\begin{lemma}\label{lem:causal_fd_levels}
Assume that \eqref{eq:causal_model_levels} and \eqref{eq:causal_model_fd} hold. Then $\forall g$ $\exists \alpha_g$ such that $\alpha_g=\alpha_{g,t}=\alpha^{fd}_{g,t}$ $\forall t$.
\end{lemma}
To avoid any confusion, note that at the same time, it follows from \eqref{eq:obs_outcome} that a causal model in levels with time-varying effects is compatible with a causal model in first-difference with time-varying effects, if the first-differenced outcome is affected both by the first-differenced treatment and by the baseline treatment. Instead, \eqref{eq:causal_model_fd} assumes that the first-differenced outcome is only affected by the first-differenced treatment. Following \eqref{eq:obs_outcome}, one may redefine the residual in \eqref{eq:causal_model_fd} as $\Delta Y_{g,t}(\tilde{0})=\Delta Y_{g,t}(0)+\Delta \alpha_{g,t}\times D_{g,t-1}$, leaving the dependence in $D_{g,t-1}$ implicit. However, with this model, all the exogeneity
conditions below (e.g. Assumption 2 or 6) have to hold with $\Delta Y_{g,t}(0)+\Delta \alpha_{g,t}\times D_{g,t-1}$ instead of $\Delta Y_{g,t}(0)$. This renders these
assumptions less plausible, because $D_{g,t-1}$ is likely to be correlated with $Z_{g,t-1}$. For instance, Assumption 2 then essentially requires that $\cov(\Delta Z_{g,t}, Z_{g,t-1}) = 0$, a strong and testable requirement (see Section \ref{sec_as_good_as_randomly_assigned} for further discussion).

\subsection{Estimator and estimand}

To simplify exposition, for now we assume that $T=2$. In Web Appendix \ref{sec_multipleperiods}, we extend some of our results to applications with multiple time periods. Accordingly, the data contains only one first-difference, and for any variable $R$, $\Delta R_{g}$ stands for $R_{g,2}-R_{g,1}$.

\begin{definition}[First-difference 2SLS estimator]\label{def_2sls_c}
Let ${\Delta Z}_{.}=\frac{1}{G} \sum_{g=1}^G \Delta Z_{g}$, and
let
	\begin{align*}
	\hat{\theta}^{b}
	&=\frac{\sum_{g=1}^G \Delta Y_{g} \left( \Delta Z_{g}-{\Delta Z}_{.}\right) }{\sum_{g=1}^G \Delta D_{g} \left( \Delta Z_{g}-{\Delta Z}_{.}\right) }.
	\end{align*}
\end{definition}
$\hat{\theta}^{b}$ is the sample coefficient from a 2SLS regression of $\Delta Y_{g}$ on an intercept and $\Delta D_{g}$, using $\Delta Z_{g}$ as the instrument. Throughout the paper, we consider unweighted regressions. In Web Appendix \ref{sec_multipleperiods}, we extend some of our decompositions to weighted regressions. We do not extend our decompositions to regressions with covariates, but doing so would be a mechanical extension.

\paragraph{Bartik instrument \citep{bartik1991benefits}.} Though several of our results apply to any FD 2SLS regression, some assume that the instrument has a shift-share structure. Assume there are $S$ sectors indexed by $s\in \{1,...,S\}$. Let $Z_{s,t}$ denote a shock affecting sector $s$ at period $t$.
\begin{definition}\label{def_bartik}
	For all $(g,t)$, the Bartik instrument $Z_{g,t}$ is:
	\begin{equation*}
	Z_{g,t}=\sum_{s=1}^S Q_{s,g} Z_{s,t}.
	\end{equation*}
	\end{definition}
For all $(g,t)$, $Q_{s,g}$ are positive weights summing to 1 or less, reflecting the importance of sector $s$ in location $g$ at period $t$. For instance, $Q_{s,g}$ could be the share that sector $s$ accounts for in $g$'s employment at $t=1$. Definition \ref{def_bartik} assumes time-invariant shares: all our results can readily be extended to allow for time-varying shares. In Bartik designs, two approaches to statistical uncertainty have been proposed. In the first one, proposed by \cite{goldsmith2020bartik} and hereafter referred to as the shares approach, the shocks $Z_{s,t}$ are conditioned upon, and locations are an independent and identically distributed (iid) sample drawn from a super population of locations. Then, the vectors $(\Delta Z_g,\Delta D_g,\Delta Y_g)$ are iid. In the second one, proposed by \cite{borusyak2020quasi} and hereafter referred to as the shocks approach, the locations are conditioned upon, and the shocks $(Z_{s,1},Z_{s,2})$ are drawn independently across sectors. 


\begin{definition}[First-difference 2SLS estimand]\label{def_2sls_c_estimand}
Let
	\begin{align}
	\theta^{b}&=\frac{\sum_{g=1}^G E\left(\Delta Y_{g} \left( \Delta Z_{g}-E\left({\Delta Z}_{.}\right)\right)\right) }{\sum_{g=1}^G E\left(\Delta D_{g} \left( \Delta Z_{g}-E\left({\Delta Z}_{.}\right)\right)\right)}.\label{eq:thm3_1}
	\end{align}
\end{definition}
In Bartik designs, in the shares approach of \cite{goldsmith2020bartik}, locations are iid, so $\theta^{b}=\cov(\Delta Y,\Delta Z)/\cov(\Delta D,\Delta Z)$, the probability limit of $\hat{\theta}^{b}$ when $G\rightarrow +\infty$. Similarly, in the shocks approach of \cite{borusyak2020quasi}, $\hat{\theta}^{b}-\theta^{b}$ converges to zero when $S\rightarrow +\infty$.

\paragraph{Instrument relevance.} Throughout the paper, we assume that the instrument is relevant: $\sum_{g=1}^G E\left(\Delta D_{g} \left( \Delta Z_{g}-E\left({\Delta Z}_{.}\right)\right)\right)\ne 0$. Without loss of generality we can further assume that $\sum_{g=1}^G E\left(\Delta D_{g} \left( \Delta Z_{g}-E\left({\Delta Z}_{.}\right)\right)\right)> 0$: the population first-stage is strictly positive.

\subsection{Definition of robustness to heterogeneous effects}\label{sub_def_rob}

Robustness to heterogeneous effects plays a key role in this paper, so we formally define the robustness concept we use.
\begin{definition}\label{def:rob_hetX}
$\theta^{b}$ is robust to heterogeneous effects if and only if $\theta^{b}=E\left(\sum_{g=1}^G \sum_{t=1}^2 w_{g,t} \alpha_{g,t}\right),$ with $E\left(\sum_{g=1}^G \sum_{t=1}^2 w_{g,t}\right)=1$ and $w_{g,t}\geq 0$ almost surely.
\end{definition}

\paragraph{Strenghtening Definition \ref{def:rob_hetX} to require that $\theta^{b}$ identifies the average treatment effect (ATE)?}
One may find Definition \ref{def:rob_hetX} too weak, and argue that $\theta^{b}$ is only robust to heterogeneous effects if $\theta^{b}=E\left(\frac{1}{2G}\sum_{g=1}^G \sum_{t=1}^2 \alpha_{g,t}\right).$ All our results below show that $\theta^{b}$ is not robust under our weaker criterion, so $\theta^{b}$ is also not robust under any stricter criterion.

\paragraph{Weakening Definition \ref{def:rob_hetX} to require that $E\left(w_{g,t}\right)\geq 0$ instead of $w_{g,t}\geq 0$?}
$w_{g,t}\geq 0$ almost surely is a strong, refutable condition, that can be ruled out whenever at least one of the realized (i.e. ex-post) weights is negative. A weakening of this condition would be to require instead $E\left(w_{g,t}\right)\geq 0$. However, whenever $w_{g,t}$ and $\alpha_{g,t}$ are correlated, as is often likely to be the case, $E\left(w_{g,t}\right)\geq 0$ is not enough to prevent a so-called sign reversal, where, say, $\alpha_{g,t}\geq 0$ almost surely for all $(g,t)$, but $\theta^{b}<0$.\footnote{For instance, if $G=1$, $w_{1,1}=-1+3X$, $w_{1,2}=1-w_1$, $\alpha_{1,1}=1-X$, and $\alpha_{1,2}=X$, where $X$ follows a Bernoulli distribution with parameter $2/3$, then $E\left(\sum_{g=1}^G \sum_{t=1}^2 w_{g,t}\alpha_{g,t}\right) =-1$.} Our stricter condition ensures that such sign reversals cannot happen, even if $w_{g,t}$ and $\alpha_{g,t}$ are correlated.

\paragraph{Weakening Definition \ref{def:rob_hetX} to require that $E(w_{g,t}|\alpha_{g,t})\geq 0$ instead of $w_{g,t}\geq 0$?}
Another potential weakening of $w_{g,t}\geq 0$ would be to require $E(w_{g,t}|\alpha_{g,t})\geq 0$ almost surely. This weaker condition is sufficient to prevent sign-reversal. However, unless the instrument is randomly- or partly-randomly assigned, it is often impossible to assess whether $E(w_{g,t}|\alpha_{g,t})\geq 0$ holds, because $\alpha_{g,t}$ is not observed, thus making it a non-refutable condition. Moreover, without any restriction on the correlation between $\alpha_{g,t}$ and $w_{g,t}$, the two conditions are observationally equivalent. For instance, if $\alpha_{g,t}=\alpha_1 1\{w_{g,t}\geq 0\}+\alpha_2 1\{w_{g,t}< 0\}$ for two distinct real numbers $\alpha_1$ and $\alpha_2$, $E(w_{g,t}|\alpha_{g,t})\geq 0$ almost surely if and only if $w_{g,t}\geq 0$ almost surely, and $\alpha_1$ and $\alpha_2$ can be chosen to rationalize $\theta^{b}$. Observational equivalence implies that when one of the realized weights is strictly negative, we cannot rule out that $E(w_{g,t}|\alpha_{g,t})\geq 0$ fails.

\paragraph{Assessing robustness to heterogeneous effects with a random or partly-random instrument.}
As our Theorem \ref{th_2sls_c_randomshocks} below shows, when one assumes that the instrument (or part of it) is randomly assigned, it may be possible to assess whether ``$E(w_{g,t}|\alpha_{g,t})\geq 0$ almost surely'' holds. Moreover, $E(w_{g,t}|\alpha_{g,t})\geq 0$ and $w_{g,t}\geq 0$ are no longer observationally equivalent in that case. Then, we recommend replacing $w_{g,t}\geq 0$ by $E(w_{g,t}|\alpha_{g,t})\geq 0$ in our robustness definition. Our Theorem \ref{th_2sls_c_randomshocks} below shows that with a random or partly-random instrument, one may have that  $\theta^{b}$ is not robust to time-varying effects, even per this weaker robustness definition.

\paragraph{Assessing robustness to heterogeneous effects without a random or partly-random instrument.}
Researchers analyzing FD 2SLS regressions are not always willing to assume that their instrument is random or partly random. For instance, in Bartik designs, the approach to instrument exogeneity proposed by \cite{goldsmith2020bartik}, which relies on a parallel trends assumption instead of random assignment, is very popular. In such instances, to assess their regression's robustness to heterogeneous effects, we recommend that researchers follow Definition \ref{def:rob_hetX} and assess whether some of the realized weights attached to their FD 2SLS regression are negative. At the same time, to account for the fact ``random'' negative weights uncorrelated to treatment effects do not lead to sign reversal, and that with random weights $\theta^{b}$ can even identify the ATE \citep[see Corollary 2 in][]{de2020two}, we also recommend that researchers assess whether weights are correlated with plausible treatment-effect proxies.

\section{FD 2SLS regressions with heterogeneous effects}\label{sec_Bartik}

\subsection{$\theta^{b}$ is not robust to heterogeneous effects under a linear model}\label{sec_baseline}

\paragraph{Decomposition of $\theta^{b}$ under Assumption \ref{as_linear_ss}.}
\begin{theorem}\label{th_2sls_c}
	Suppose Assumption \ref{as_linear_ss} holds.
\begin{enumerate}
\item Then,
\begin{align*}
			\theta^{b}=&\frac{\sum_{g=1}^G E\left(\Delta Y_{g}(0) \left( \Delta Z_{g}-E\left({\Delta Z}_{.}\right)\right)\right) }{\sum_{g=1}^G E\left(\Delta D_{g} \left( \Delta Z_{g}-E\left({\Delta Z}_{.}\right)\right)\right)}\\
+&E \left( \sum_{g=1}^G \sum_{t=1}^2 \frac{(1\{t=2\}-1\{t=1\})D_{g,t}(\Delta Z_{g}-E\left({\Delta Z}_{.}\right))}{ E\left(  \sum_{g'=1}^G \sum_{t'=1}^2 (1\{t'=2\}-1\{t'=1\})D_{g',t'}(\Delta Z_{g'}-E\left({\Delta Z}_{.}\right)) \right)}\alpha_{g,t}\right).
\end{align*}
\item If one further assumes that for all $g$, there exists $\alpha_{g}$ such that $\alpha_{g,1}=\alpha_{g,2}=\alpha_{g}$,
\begin{align*}
			\theta^{b}=\frac{\sum_{g=1}^G E\left(\Delta Y_{g}(0) \left( \Delta Z_{g}-E\left({\Delta Z}_{.}\right)\right)\right) }{\sum_{g=1}^G E\left(\Delta D_{g} \left( \Delta Z_{g}-E\left({\Delta Z}_{.}\right)\right)\right)}+E \left( \sum_{g=1}^G \frac{ \Delta D_g (\Delta Z_{g}-E\left({\Delta Z}_{.}\right))  }{ E\left(  \sum_{g'=1}^G \Delta D_{g'}  (\Delta Z_{g'}-E\left({\Delta Z}_{.}\right)) \right)}\alpha_{g}\right).
\end{align*}
\end{enumerate}
\end{theorem}
\paragraph{Consequences of Theorem \ref{th_2sls_c}.}
Point 1 of Theorem \ref{th_2sls_c} shows that under Assumption \ref{as_linear_ss}, $\theta^{b}$ can be decomposed into the sum of two terms. The first is the population coefficient one would get from a 2SLS regression of $\Delta Y_{g} (0)$, locations' outcome evolution without treatment, on $\Delta D_{g}$, using $\Delta Z_{g}$ as the instrument. The second is the expectation of a weighted sum of the treatment effects $\alpha_{g,t}$, with weights
\begin{equation}\label{eq:weights_2SLS0}
\frac{(1\{t=2\}-1\{t=1\})D_{g,t}(\Delta Z_{g}-E\left({\Delta Z}_{.}\right))}{ E\left(  \sum_{g'=1}^G \sum_{t'=1}^2 (1\{t'=2\}-1\{t'=1\})D_{g',t'}(\Delta Z_{g'}-E\left({\Delta Z}_{.}\right)) \right)}.
\end{equation}
If $D_{g,t}>0$ for all $(g,t)$, as is for instance the case in ADH, then every location whose effects do not receive a weight equal to zero is such that either $\alpha_{g,1}$ or $\alpha_{g,2}$ is weighted negatively. Thus, exactly a half of the effects $\alpha_{g,t}$ are weighted negatively, so $\theta^{b}$ is not robust to heterogeneous effects according to our definition. Point 2 of Theorem \ref{th_2sls_c} shows that even if one assumes homogeneous effects over time, $\theta^{b}$ may still not be robust. This shows that without making further assumptions, $\theta^{b}$'s robustness does not depend on whether one posits a causal model in levels or in first-difference: with homogeneous effects over time, our causal model in levels implies a causal model in first-difference, as \eqref{eq:obs_outcome2} shows, and yet  $\theta^{b}$ may still not be robust.

\paragraph{Decomposition of $\theta^{b}$ under Assumption \ref{as_linear_ss} and an exogeneity assumption.}
\begin{assumption}\label{as_commontrend_y} (Exogenous instrument)
\begin{enumerate}
\item For all $g \in \{1,...,G\}$, $\cov(\Delta Z_{g},\Delta Y_g (0))=0$.
\item $E\left(\Delta Z_{g}\right)$ does not depend on $g$.
\end{enumerate}
\end{assumption}
Assumption \ref{as_commontrend_y} ensures that the first term in the decompositions of $\theta^{b}$ in Theorem \ref{th_2sls_c} is equal to zero. Then, it directly follows from, say, Point 1 of Theorem \ref{th_2sls_c} that under Assumptions \ref{as_linear_ss} and Assumption \ref{as_commontrend_y}, $\theta^{b}$ is equal to the weighted sum of treatment effects therein.
The first point of Assumption \ref{as_commontrend_y} requires that location $g$'s potential outcome evolution without any treatment be uncorrelated with its first-differenced instrument. This condition may be interpreted as a parallel trends assumption. The second point of Assumption \ref{as_commontrend_y} requires that $E\left(\Delta Z_{g}\right)$ does not vary across locations. In Bartik designs, Assumption \ref{as_commontrend_y} nests both the ``shares'' and ``shocks'' rationalizations of Bartik exogeneity proposed by \cite{goldsmith2020bartik} and \cite{borusyak2020quasi} and \cite{adao2019shift}. \cite{goldsmith2020bartik} consider shocks as non-stochastic, and their Assumption 2 requires that $\cov(Q_{s,g},\Delta Y_g (0))=0$. This implies Point 1 of Assumption \ref{as_commontrend_y}. Point 2 trivially holds in their setting, because they assume iid locations. In our panel data setting, with period fixed effects and no other control variables, Assumption 4.ii) in \cite{adao2019shift} requires that for all $(s,t)$, $$E\left(Z_{s,t}|\left( Y_{g,t'}(0),(Q_{s',g,t'})_{s'\in \{1,...,S\}}\right)_{(g,t')\in \{1,...,G\}\times \{1,2\}}\right)=m_t$$
for some real number $m_t$.
When shares sum to one, this implies that $E\left(\Delta Z_{g}\right)=m_2-m_1$. Then,
\begin{align*}
&\cov(\Delta Z_{g},\Delta Y_{g}(0))\\
=&E\left(\Delta Y_{g}(0)\left(\sum_{s=1}^S Q_{s,g} E\left(Z_{s,2}|\Delta Y_{g}(0),(Q_{s,g})_{s\in \{1,...,S\}}\right)\right.\right.\\
-&\left.\left.\sum_{s=1}^S Q_{s,g} E\left(Z_{s,1}|\Delta Y_{g}(0),(Q_{s,g})_{s\in \{1,...,S\}}\right)\right)\right)-(m_2-m_1)E(\Delta Y_{g}(0))\\
=&0,
\end{align*}
so Assumption \ref{as_commontrend_y} holds. In their Appendix A.1, \cite{borusyak2020quasi} allow for heterogeneous effects and also make an assumption that implies Assumption \ref{as_commontrend_y}.

\paragraph{Pretrends test of Assumption \ref{as_commontrend_y}.} Assumption \ref{as_commontrend_y} is ``placebo testable'', when the data contains prior periods where all locations are untreated, as is sometimes the case. Then, locations' outcome evolutions without any treatment are observed at those periods, and one can assess if those evolutions are correlated with locations' first-differenced instrument.


\paragraph{Connection with previous literature.} When $\Delta Z_g=\Delta D_g$, meaning that $\hat{\theta}^{b}$ is actually an OLS regression coefficient, the weights in Point 1 of Theorem \ref{th_2sls_c} reduce to those in the decomposition of OLS TWFE regressions under a parallel trends assumption in Theorem 1 of \cite{de2020two}, in the special case where $T=2$. Thus, Point 1 of Theorem \ref{th_2sls_c} may be seen as a generalization of that result to 2SLS regressions, in the special case where $T=2$. \cite{de2020two} do not give the closed-form expression of the weights in their decomposition in the special case where $T=2$. That closed-form expression can readily be obtained from Point 1 of Theorem \ref{th_2sls_c},     replacing $\Delta Z_g$ by $\Delta D_g$, and it might be of independent interest. For instance, it follows from Point 1 of Theorem \ref{th_2sls_c} that when $T=2$ and $D_{g,t}>0$ for all $(g,t)$, exactly 50\% of the non-zero weights attached to OLS TWFE regressions are negative, a fact that was not noted in \cite{de2020two}. With iid locations, Point 2 of Theorem \ref{th_2sls_c} reduces to
\begin{align*}
			\theta^{b}=E \left(\frac{\Delta D (\Delta Z-E(\Delta Z))}{ E\left(\Delta D (\Delta Z-E(\Delta Z))\right)}\alpha\right),
\end{align*}
a first-difference version of a known result for cross-sectional IV regressions under a linear treatment effect model \citep[see e.g. Equation (3) in][]{benson2022ivcrc}. Point 2 of Theorem \ref{th_2sls_c} shows that a similar result holds in first-difference if the treatment effect is constant over time, as then one has a linear treatment effect model in first-difference, as shown in Equation \eqref{eq:obs_outcome2}. In the cross-sectional case, the numerator of the weights is $D (Z-E(Z))$. As $D$ is positive, weights are strictly negative if and only if $D>0$ and $Z<E(Z)$. In the panel case,  $\Delta D$ may be negative, so weights are strictly negative if and only if $\Delta D$ and $\Delta Z-E(\Delta Z)$ are different from zero and of a different sign, thus leading to a different characterization of the negatively-weighted effects.

\subsection{In Bartik designs, $\theta^{b}$ is still not robust if one assumes a linear first-stage}\label{sec_firststage}

Throughout this section and the next, we assume that the instrument satisfies Definition \ref{def_bartik}: we are in a Bartik design, with a shift-share instrument.

\paragraph{Linear first-stage model.} For any $(z_1,...,z_S)\in \mathbb{R}^S$, let $D_{g,t}(z_1,...,z_S)$ denote the potential treatment of location $g$ at period $t$ if $(Z_{1,t},...,Z_{S,t})=(z_1,...,z_S)$.
And let $D_{g,t}(\bm{0})=D_{g,t}(0,...,0)$ denote the potential treatment of $g$ at $t$ without any shocks. The actual treatment of $g$ at $t$ is $D_{g,t}=D_{g,t}(Z_{1,t},...,Z_{S,t}).$ We make the following assumption:
	\begin{assumption}\label{as_linear_fs}
		Linear First-Stage Model: for all $(g,t) \in \{1,...,G\}\times \{1,...,T\}$, there exists $(\beta_{s,g,t})_{s\in \{1,...,S\}}$ such that for any $(z_1,...,z_S)$:
		\begin{align*}
		D_{g,t} (z_1,...,z_s) = D_{g,t} (\bm{0}) + \sum_{s=1}^S Q_{s,g}  \beta_{s,g,t}z_s.
		\end{align*}
	\end{assumption}
Assumption \ref{as_linear_fs} requires that the effect of the shocks on the treatment be linear: increasing $Z_{s,t}$ by 1 unit, holding all other shocks constant, increases the treatment of location $g$ at period $t$ by $Q_{s,g} \beta_{s,g,t}$ units. Similar assumptions are also made by \citet{adao2019shift} (see their Equation (11)) and \citet{goldsmith2020bartik} (see their Equation (8), which we discuss in more details later). Under Assumption \ref{as_linear_fs},
		\begin{align}\label{eq:obs_treat}
		D_{g,t} = D_{g,t} (\bm{0}) + \sum_{s=1}^S Q_{s,g}\beta_{s,g,t}Z_{s,t}.
		\end{align}
Note that if $\beta_{s,g,t}=\beta_{g,t}$,
\begin{align}\label{eq:obs_treat}
		D_{g,t} = D_{g,t} (\bm{0}) + \beta_{g,t}Z_{g,t},
		\end{align}
a first-stage model that only depends on the instrument $Z_{g,t}$, not on the shocks. Thus, while Theorem \ref{th_2sls_c_fs} applies to FD 2SLS Bartik regressions, it can also be used to derive decompositions of any FD 2SLS regression under a linear first-stage model in the instrument and Assumption \ref{as_commontrend_rfy}, replacing $\beta_{s,g,t}$ by $\beta_{g,t}$. Note also that if the first-stage effects are constant over time ($\beta_{s,g,2}=\beta_{s,g,1}$ for all $g$), \eqref{eq:obs_treat} implies
\begin{align}\label{eq:obs_treat2}
		\Delta D_{g} = \Delta D_{g} (\bm{0}) + \sum_{s=1}^S Q_{s,g}\beta_{s,g}\Delta Z_{s},
		\end{align}
a linear first-stage model relating the first-differenced treatment and shocks. With a slight abuse of notation, let
$$\Delta Y_g (D_{g}(\bm{0}))=Y_{g,2}(0)+\alpha_{g,2}D_{g,2} (\bm{0})-(Y_{g,1}(0)+\alpha_{g,1}D_{g,1} (\bm{0}))$$
denote the outcome evolution that location $g$ would have experienced from period one to two without any shocks. Plugging \eqref{eq:obs_treat} into \eqref{eq:obs_outcome} yields the following first-differenced reduced-form equation:
\begin{align}\label{eq:obs_outcome3}
		\Delta Y_{g} = \Delta Y_g (D_{g}(\bm{0})) + \alpha_{g,2} \sum_{s=1}^S Q_{s,g}\beta_{s,g,2}Z_{s,2}-\alpha_{g,1} \sum_{s=1}^S Q_{s,g}\beta_{s,g,1}Z_{s,1}.
		\end{align}
If the first-stage and treatment effects are constant over time, \eqref{eq:obs_outcome3} implies
\begin{align}\label{eq:obs_outcome4}
		\Delta Y_{g} = \Delta Y_g (D_{g}(\bm{0})) + \alpha_{g}\sum_{s=1}^S Q_{s,g}\beta_{s,g}\Delta Z_{s}.
		\end{align}

\paragraph{Identifying assumption with a first-stage model.}
With our first-stage model in hand, the identifying assumption we consider requires that the instrument be uncorrelated with the reduced-form and first-stage residuals $\Delta Y_g (D_{g}(\bm{0}))$ and $\Delta D_{g} (\bm{0})$, rather than with the second-stage residual $\Delta Y_g (0)$.
\begin{assumption}\label{as_commontrend_rfy} (Exogenous instrument, v2)
\begin{enumerate}
\item For all $g \in \{1,...,G\}$, $\cov(\Delta Z_{g},\Delta Y_g (D_{g} (\bm{0})))=0$.
\item For all $g \in \{1,...,G\}$, $\cov(\Delta Z_{g},\Delta D_{g} (\bm{0}))=0$.
\item $E\left(\Delta Z_{g}\right)$ does not depend on $g$.
\end{enumerate}
\end{assumption}
Assumption \ref{as_commontrend_rfy} is similar to the parallel trends conditions considered by \cite{de2013note} and \cite{hudson2017interpreting}.
The random-shocks assumption in \cite{borusyak2020quasi} and \cite{adao2019shift} implies Assumption \ref{as_commontrend_rfy}. Assuming $\cov(Q_{s,g},\Delta Y_g (D_{g} (\bm{0})))=0$, $\cov(Q_{s,g},\Delta D_{g} (\bm{0}))=0$, non-stochastic shocks, and iid locations, in the spirit of \cite{goldsmith2020bartik}, also implies Assumption \ref{as_commontrend_rfy}.

\paragraph{Comparing Assumptions \ref{as_commontrend_y} and \ref{as_commontrend_rfy}.} If $E(\Delta D_{g}(\bm{0}))=0$ and $\alpha_{g,1}=\alpha_{g,2}=\alpha_{g}$, Assumptions \ref{as_commontrend_y} and \ref{as_commontrend_rfy} can jointly hold under no restrictions on the joint distribution of $\alpha_g$ and $\Delta Z_{g}$. For instance, if Point 1 of Assumption \ref{as_commontrend_y} holds and $E(\Delta D_{g}(\bm{0})|\alpha_g,\Delta Z_{g})=0$, then Points 1 and 2 of Assumption \ref{as_commontrend_rfy} hold. On the other hand, if $E(\Delta D_{g}(\bm{0}))\ne 0$ or $\alpha_{g,1}\ne \alpha_{g,2}$, imposing jointly Assumptions \ref{as_commontrend_y} and \ref{as_commontrend_rfy} is essentially equivalent to assuming that $\cov(\Delta Z_{g},\alpha_{g,1})=\cov(\Delta Z_{g},\alpha_{g,2})=0$, a strong requirement, unless one is ready to assume that the first-differenced instrument is randomly assigned to locations. Our decompositions of $\theta^{b}$ under Assumption \ref{as_commontrend_rfy} in Theorem \ref{th_2sls_c_fs} below are similar to those under Assumption \ref{as_commontrend_y} that follow from Theorem \ref{th_2sls_c}. Imposing Assumption \ref{as_commontrend_y} or \ref{as_commontrend_rfy} does not change much our assessment of $\theta^{b}$'s robustness to heterogenous effects.

\paragraph{Decompositions of $\theta^{b}$ under Assumptions \ref{as_linear_ss} and \ref{as_linear_fs}-\ref{as_commontrend_rfy}.}
\begin{theorem}\label{th_2sls_c_fs}
	Suppose the instrument satisfies Definition \ref{def_bartik}, and Assumptions \ref{as_linear_ss} and \ref{as_linear_fs}-\ref{as_commontrend_rfy} hold.
	\begin{enumerate}
    \item Then,
    \begin{adjustwidth}{-50pt}{-50pt}
		\begin{align*}
			\theta^{b}=E \left( \sum_{g=1}^G \sum_{t=1}^2   \frac{(1\{t=2\}-1\{t=1\})\sum_{s=1}^S Q_{s,g} \beta_{s,g,t}Z_{s,t} (\Delta Z_{g}-E\left({\Delta Z}_{.}\right)) }{ E\left(  \sum_{g'=1}^G \sum_{t'=1}^2   (1\{t'=2\}-1\{t'=1\})\sum_{s=1}^S Q_{s,g'} \beta_{s,g',t'}Z_{s,t'} (\Delta Z_{g'}-E\left({\Delta Z}_{.}\right))\right)  }\alpha_{g,t} \right).
		\end{align*}
\end{adjustwidth}
		\item If one further assumes that for all $g$, there exist $\alpha_{g}$ and $(\beta_{s,g})_{s\in\{1,...,S\}}$ such that $\alpha_{g,1}=\alpha_{g,2}=\alpha_{g}$ and $\beta_{s,g,1}=\beta_{s,g,2}=\beta_{s,g}$, then
		\begin{align*}
			\theta^{b}=E \left( \sum_{g=1}^G   \frac{ \sum_{s=1}^S Q_{s,g}\beta_{s,g}\Delta Z_s (\Delta Z_{g}-E\left({\Delta Z}_{.}\right))}{ E\left(  \sum_{g'=1}^G \sum_{s=1}^S Q_{s,g'} \beta_{s,g'}\Delta Z_{s}  (\Delta Z_{g'}-E\left({\Delta Z}_{.}\right)) \right)  }\alpha_{g} \right).
		\end{align*}
\item If on top of the assumptions in Point 2, one further assumes that $\beta_{s,g}=\beta$,
		\begin{align*}
			\theta^{b}=E \left( \sum_{g=1}^G \frac{ \Delta Z_g (\Delta Z_{g}-E\left({\Delta Z}_{.}\right))  }{ E\left(  \sum_{g'=1}^G \Delta Z_{g'}  (\Delta Z_{g'}-E\left({\Delta Z}_{.}\right)) \right) }\alpha_{g}\right).
		\end{align*}
	\end{enumerate}
\end{theorem}

\paragraph{Consequences of Theorem \ref{th_2sls_c_fs}} Point 1 of Theorem \ref{th_2sls_c_fs} shows that under Assumptions \ref{as_linear_ss} and \ref{as_linear_fs}-\ref{as_commontrend_rfy}, $\theta^{b}$ identifies a weighted sum of the treatment effects $\alpha_{g,t}$, with weights
\begin{equation}\label{eq:weights_2SLS_fs}
 \frac{(1\{t=2\}-1\{t=1\})\sum_{s=1}^S Q_{s,g} \beta_{s,g,t}Z_{s,t} (\Delta Z_{g}-E\left({\Delta Z}_{.}\right)) }{ E\left(  \sum_{g'=1}^G \sum_{t'=1}^2   (1\{t'=2\}-1\{t'=1\})\sum_{s=1}^S Q_{s,g'} \beta_{s,g',t'}Z_{s,t'} (\Delta Z_{g'}-E\left({\Delta Z}_{.}\right))\right)  }.
\end{equation}
Those weights are identical to those in \eqref{eq:weights_2SLS0}, replacing $D_{g,t}$ by $\sum_{s=1}^S Q_{s,g} \beta_{s,g,t}Z_{s,t}$, the effect of the shocks on $D_{g,t}$. Therefore, unlike the weights in \eqref{eq:weights_2SLS0}, those in \eqref{eq:weights_2SLS_fs} cannot be estimated, as they depend on the first stage effects $\beta_{s,g,t}$. Let us assume that $Z_{s,t}>0$ for all $(s,t)$, as is  for instance the case in ADH. If one further assumes that the first-stage effects $\beta_{s,g,t}$ are all positive, an assumption similar to the monotonicity condition in \cite{imbens1994identification}, then every location is such that either $\alpha_{g,1}$ or $\alpha_{g,2}$ is weighted negatively. Therefore, adding a linear first-stage model with a monotonicity condition is not enough to make $\theta^{b}$ robust to heterogeneous effects.  Point 2 of Theorem \ref{th_2sls_c_fs} shows that even assuming that the first-stage and treatment effects are homogeneous over time, $\theta^{b}$ may still not be robust to heterogeneous effects across locations. Finally, Point 3 shows that even if one further assumes a fully homogeneous first-stage effect, $\theta^{b}$ may still not be robust. The weights in that last decomposition can be estimated.

\paragraph{Comparing Point 2 of Theorem \ref{th_2sls_c_fs} to Equation (10) in \cite{goldsmith2020bartik}.} In their Equation (10), \cite{goldsmith2020bartik} analyze a Bartik regression with one time period, in a model with location-specific treatment effects (see their Equation (7)). The regression they consider nests that in our Definition \ref{def_2sls_c}, if the treatment and outcome in their regression are first-differenced. Then, their Equation (7) is a linear model in first-difference, which assumes constant effects over time, so their Equation (10) should be compared to Point 2 of our Theorem \ref{th_2sls_c_fs}. Like Point 2 of our Theorem \ref{th_2sls_c_fs}, their Equation (10) shows that $\theta^{b}$ identifies a weighted sum of treatment effects, potentially with some negative weights. However, the weights in their and our decomposition differ. Expressed in our notation, the weight assigned to $\alpha_g$ in their decomposition is
\begin{align*}
&\frac{\sum_{s=1}^S\left(\Delta Z_s \left(\sum_{g'=1}^G Q_{s,g'}(\Delta D_{g'}-\Delta D_{.})\right)\left(Q_{s,g}-Q_{s,.}\right)^2\Delta Z_s \beta_{s,g}\right)}{\left(\sum_{s=1}^S \Delta Z_s \left(\sum_{g'=1}^G Q_{s,g'}(\Delta D_{g'}-\Delta D_{.})\right)\right)\times \left(\sum_{g=1}^G\left(Q_{s,g}-Q_{s,.}\right)^2\Delta Z_s \beta_{s,g}\right)},\nonumber
\end{align*}
where $Q_{s,.}=\frac{1}{G}\sum_{g=1}^G Q_{s,g}$ is the average share of sector $s$ across locations, and where
$$\frac{\Delta Z_s \left(\sum_{g'=1}^G Q_{s,g'}(\Delta D_{g'}-\Delta D_{.})\right)}{\sum_{s=1}^S \Delta Z_s \left(\sum_{g'=1}^G Q_{s,g'}(\Delta D_{g'}-\Delta D_{.})\right)}$$
is the so-called Rotemberg weight (see \citealt{rotemberg1983instrument}). The weights in our decomposition do not depend on the Rotemberg weights.

\paragraph{Why do Point 2 of Theorem \ref{th_2sls_c_fs} and Equation (10) in \cite{goldsmith2020bartik} differ?}
The difference between our decompositions stems from the fact our first-stage assumptions are different and almost incompatible. In what follows, we assume that shocks are non-stochastic, as in \citet{goldsmith2020bartik}.
Then, using our notation, and assuming the regression has no control variables, the first-stage assumptions in \citet{goldsmith2020bartik} (see Equation (8) and Assumption 3 therein) require that for all $(s,g)$,
\begin{align}
&\Delta D_g=\mu^D+Q_{s,g}\Delta Z_s\beta_{s,g}+u_{s,g},\label{eq:FS_GSS}\\
&\text{with }E(Q_{s,g}u_{s,g}\alpha_g)=0.\label{eq:orth_GSS}
\end{align}
\eqref{eq:FS_GSS} is a first-differenced first-stage model similar to \eqref{eq:obs_treat2}, where the effect of only one sector appears explicitly. \eqref{eq:obs_treat2} and \eqref{eq:FS_GSS} imply that $u_{s,g}=\Delta D_{g}(\bm{0})-\mu^D+\sum_{s'\ne s}Q_{s',g}\Delta Z_{s'}\beta_{s',g},$
so
\begin{align*}
E(Q_{s,g}u_{s,g}\alpha_g)=&E(Q_{s,g}(\Delta D_{g}(\bm{0})-\mu^D)\alpha_g)+\sum_{s'\ne s}E(Q_{s,g}Q_{s',g}\alpha_g \beta_{s',g})\Delta Z_{s'}.
\end{align*}
Then, \eqref{eq:orth_GSS} is hard to rationalize. For instance, if for all $(s,g)$ $\Delta Z_{s}>0$, $\beta_{s,g}>0$, $\alpha_g>0$, and $Q_{s,g}>0$,
$\sum_{s'\ne s}E(Q_{s,g}Q_{s',g}\alpha_g \beta_{s',g})\Delta Z_{s'}>0$, so \eqref{eq:orth_GSS} can only hold if the first and second terms in the right-hand-side of the previous display cancel each other out.
Overall, whenever the linear first-stage model with time-invariant effects in \eqref{eq:obs_treat2} seems plausible, the first-stage assumptions in \cite{goldsmith2020bartik} are unlikely to hold, and the decomposition of $\theta^{b}$ in their Equation (10) is also unlikely to hold. Heterogeneous effects is not a central issue in \citet{goldsmith2020bartik}. Except for their Equation (10), all their other results assume homogeneous effects and do not rest on their Equation (8) and Assumption 3.

\subsection{In Bartik designs, $\theta^{b}$ may still not be robust with randomly-assigned shocks}\label{sec_as_good_as_randomly_assigned}

\paragraph{The random-shocks assumption.}
Let $\mathcal{F}=\left(Y_{g,t}(0),D_{g,t}(\bm{0}),\alpha_{g,t},(Q_{s,g},\beta_{s,g,t})_{s\in \{1,...,S\}}\right)_{(g,t)\in \{1,...,G\}\times \{1,2\}}$.
\begin{assumption}(Random shocks) \label{as_randomly_assigned_shocks}
\begin{enumerate}
\item For all $(s,t)$, $E\left(Z_{s,t}|\mathcal{F}\right)=E\left(Z_{s,t}\right)$.
\item For all $t$, there exists a real number $m_t$ such that $E\left(Z_{s,t}\right)=m_t$ for all $s$.
\item The vectors $(Z_{s,1},Z_{s,2})$ are mutually independent across $s$, conditional on $\mathcal{F}$.
\end{enumerate}
\end{assumption}
Point 1 of Assumption \ref{as_randomly_assigned_shocks} requires that shocks be mean independent of locations' potential outcomes without treatment, potential treatments without shocks, shares, and first-stage and treatment effects. Point 2 requires that at every period, all sector-level shocks have the same expectation.
Point 3 requires that the vector of period-one and period-two shocks be independent across sectors, but it allows for serial correlation within sectors. Points 1 and 2 of Assumption \ref{as_randomly_assigned_shocks} are equivalent to Assumption 4.ii) in \cite{adao2019shift} with panel data, period fixed effects, and no other control variables. Point 3 is identical to the independence assumption that \cite{adao2019shift} make in their Section V.A, with panel data and clusters defined as sectors.

\paragraph{Decomposition of $\theta^{b}$ under Assumptions \ref{as_linear_ss}, \ref{as_linear_fs}, and \ref{as_randomly_assigned_shocks}.}
\begin{theorem}\label{th_2sls_c_randomshocks}
Suppose the instrument satisfies Definition \ref{def_bartik}, Assumptions \ref{as_linear_ss}, \ref{as_linear_fs}, and  \ref{as_randomly_assigned_shocks} hold, and $\sum_{s=1}^S Q_{s,g}=1$ for all $(g,t)$. If one also assumes that for all $s$ and $(t,t')\in \{1,2\}^2$, $E\left(Z_{s,t}Z_{s,t'}|\mathcal{F}\right)=E\left(Z_{s,t}Z_{s,t'}\right)$, then
\begin{adjustwidth}{-50pt}{-50pt}
\begin{align}\label{eq:resultAKM}
\theta^{b}&=E\left(\sum_{g=1}^G \sum_{t=1}^2   \frac{\sum_{s=1}^S\beta_{s,g,t}Q^2_{s,g}\left(V\left(Z_{s,t}\right)-\cov\left(Z_{s,1},Z_{s,2}\right)\right)}{E\left(\sum_{g'=1}^G \sum_{t'=1}^2 \sum_{s=1}^S\beta_{s,g',t'}Q^2_{s,g'}\left(V\left(Z_{s,t'}\right)-\cov\left(Z_{s,1},Z_{s,2}\right)\right)\right)} \alpha_{g,t}\right).
\end{align}
\end{adjustwidth}
\end{theorem}
\paragraph{Remarks on the assumptions underlying Theorem \ref{th_2sls_c_randomshocks}.} On top of Assumption \ref{as_randomly_assigned_shocks}, Theorem \ref{th_2sls_c_randomshocks} further assumes that shares sum to one. If that is not the case, \cite{borusyak2020quasi} show that one should not estimate $\theta^{b}$ under their random-shocks assumption. Instead, one should replace the intercept by locations' sum of shares in the FD 2SLS Bartik regression. We conjecture that when shares do not sum to one, a result similar to that in Theorem \ref{th_2sls_c_randomshocks} can be shown for that estimand. Theorem \ref{th_2sls_c_randomshocks} also further assumes that $E\left(Z_{s,t}Z_{s,t'}|\mathcal{F}\right)=E\left(Z_{s,t}Z_{s,t'}\right)$, a mild strengthening of Point 1 of Assumption  \ref{as_randomly_assigned_shocks}.

\paragraph{The weights in Theorem \ref{th_2sls_c_randomshocks} are all positive if $\cov\left(Z_{s,1},Z_{s,2}\right)\leq 0$ for all $s$, or if $V\left(Z_{s,1}\right)=V\left(Z_{s,2}\right)$ for all $s$.} However, there are applications where those two conditions are violated. For instance, in the data of ADH, we find that the sample variance of $Z_{s,2}$ is more than 3 times larger than the sample variance of $Z_{s,1}$ (imports from China are strongly increasing over the study period), while the sample correlation of $Z_{s,1}$ and $Z_{s,2}$ is equal to 0.70.\footnote{There are three periods in ADH. The numbers in the text are computed for the first two periods in their data. Results are similar if one instead uses the last two periods.} 

\paragraph{The weights in Theorem \ref{th_2sls_c_randomshocks} are also all positive if $\cov\left(Z_{s,1},Z_{s,2}-Z_{s,1}\right)=0$.}
In that case, Theorem \ref{th_2sls_c_randomshocks} simplifies to
\begin{align}\label{eq:resultAKM2}
\theta^{b}&=E\left(\sum_{g=1}^G \frac{\sum_{s=1}^S\beta_{s,g,t}Q^2_{s,g}V\left(Z_{s,2}-Z_{s,1}\right)}{E\left(\sum_{g'=1}^G \sum_{s=1}^S\beta_{s,g',2}Q^2_{s,g'}V\left(Z_{s,2}-Z_{s,1}\right)\right)} \alpha_{g,2}\right).
\end{align}
so $\theta^{b}$ is robust to heterogeneous effects. However, $\cov\left(Z_{s,1},Z_{s,2}-Z_{s,1}\right)=0$ is a strong, testable requirement, and there are applications where this condition is strongly violated. For instance, in the data of ADH, we find that the sample correlation between $Z_{s,1}$ and $Z_{s,2}-Z_{s,1}$ is equal to $0.41$. If $Z_{s,2}$ and $Z_{s,1}$ have the same support, $Z_{s,2}-Z_{s,1}$ and $Z_{s,1}$ cannot be independent, thus making it unlikely, and sometimes impossible\footnote{For instance, $\cov\left(Z_{s,1},Z_{s,2}-Z_{s,1}\right)<0$ if $Z_{s,1}$ and $Z_{s,2}$ are identically distributed and not perfectly correlated Bernoulli variables.} that they are uncorrelated. Relatedly, in their Equation (6), \cite{borusyak2023} give a sufficient condition to have only positive weights in our Theorem \ref{th_2sls_c_randomshocks}, which requires that $Z_{s,2}-Z_{s,1}$ be uncorrelated with a residual that depends on $D_{g,1}$. If $D_{g,1}$ is caused by or at least correlated with $Z_{s,1}$, their orthogonality condition may be hard to rationalize without assuming $\cov\left(Z_{s,1},Z_{s,2}-Z_{s,1}\right)=0$.

\paragraph{The weights in Theorem \ref{th_2sls_c_randomshocks} are also all positive if the first-stage and treatment effects do not change over time.} Indeed, if $\beta_{s,g,t}=\beta_{s,g}$ and $\alpha_{g,t}=\alpha_{g}$,
Theorem \ref{th_2sls_c_randomshocks} simplifies to
\begin{align}\label{eq:resultAKM3}
\theta^{b}&=E\left(\sum_{g=1}^G \frac{\sum_{s=1}^S\beta_{s,g}Q^2_{s,g}V\left(Z_{s,2}-Z_{s,1}\right)}{E\left(\sum_{g'=1}^G \sum_{s=1}^S\beta_{s,g'}Q^2_{s,g'}V\left(Z_{s,2}-Z_{s,1}\right)\right)} \alpha_{g}\right).
\end{align}

\paragraph{Outside of those special cases, some of the weights in Theorem \ref{th_2sls_c_randomshocks} may be negative.}
\eqref{eq:resultAKM} shows that under our causal model in levels in Assumption \ref{as_linear_ss}, and outside of the aforementioned special cases, $\theta^{b}$ may not be robust to heterogeneous effects, even with randomly-assigned shocks. Note that $V\left(Z_{s,t}\right)-\cov\left(Z_{s,1},Z_{s,2}\right)$, the potentially negative quantity in the weights, is non-random. Therefore, the expectation of the weights conditional on $\alpha_{g,t}$ can also be negative. Thus, $\theta^{b}$ may not be robust, even with random shocks and under the weaker robustness definition discussed in Section \ref{sub_def_rob}. Let us further assume that shocks' second moments do not depend on $s$: $V\left(Z_{s,t}\right)=\sigma^2_t$ and $\cov\left(Z_{s,1},Z_{s,2}\right)=\rho \sigma_1 \sigma_{2}$, an assumption in the spirit of Point 2 of Assumption \ref{as_randomly_assigned_shocks}. Then, the weights in \eqref{eq:resultAKM} simplify to
$$\frac{\left(\sigma^2_t-\rho \sigma_1 \sigma_{2}\right)\sum_{s=1}^S\beta_{s,g,t}Q^2_{s,g}}{E\left(\sum_{g'=1}^G \sum_{t'=1}^2 \left(\sigma^2_{t'}-\rho \sigma_1 \sigma_{2}\right)\sum_{s=1}^S\beta_{s,g',t'}Q^2_{s,g'}\right)}.$$
If $\beta_{s,g,t}\geq 0$ for all $(s,g,t)$, the weights are of the same sign as $\sigma^2_t-\rho \sigma_1 \sigma_{2}$, which can be estimated ($\rho$ is just the correlation between the period-one and period-two shock of the same sector). In ADH, $\hat{\sigma}^2_1-\hat{\rho} \hat{\sigma}_1 \hat{\sigma}_{2}<0$, so the estimated weight on $\alpha_{g,1}$ is negative for all $g$.

\paragraph{Results similar to Theorem \ref{th_2sls_c_randomshocks} apply to FD 2SLS (resp. OLS) regressions with a random instrument (resp. treatment).}
Letting $S=1$, it follows from Theorem \ref{th_2sls_c_randomshocks} that with as-good-as randomly assigned instruments $(Z_{g,1},Z_{g,2})$ (which, in a Bartik design, is stronger than assuming as-good-as randomly assigned shocks), we have that for any FD 2SLS regression,
\begin{align*}
\theta^{b}&=E\left(\sum_{g=1}^G \sum_{t=1}^2   \frac{\beta_{g,t}\left(V\left(Z_{g,t}\right)-\cov\left(Z_{g,1},Z_{g,2}\right)\right)}{E\left(\sum_{g'=1}^G \sum_{t'=1}^2\beta_{g',t'}\left(V\left(Z_{g',t'}\right)-\cov\left(Z_{g',1},Z_{g',2}\right)\right)\right)} \alpha_{g,t}\right).
\end{align*}
Then, replacing the instrument by the treatment, it follows that for any FD OLS regression with as-good-as randomly assigned treatments $(D_{g,1},D_{g,2})$, the treatment coefficient is equal to
\begin{align*}
&E\left(\sum_{g=1}^G \sum_{t=1}^2   \frac{V\left(D_{g,t}\right)-\cov\left(D_{g,1},D_{g,2}\right)}{E\left(\sum_{g'=1}^G \sum_{t'=1}^2V\left(D_{g',t'}\right)-\cov\left(D_{g',1},D_{g',2}\right)\right)} \alpha_{g,t}\right).
\end{align*}
The weights in the previous display are guaranteed to be positive if $D_{g,t}$ is binary, but not otherwise. Thus, the negative weights in Theorem \ref{th_2sls_c_randomshocks} are not specific to Bartik regressions. Rather, they arise from first-differencing.\footnote{We are grateful to Peter Hull for noting this point.} It has been shown that with a binary randomized treatment, OLS TWFE regressions always estimate a convex combination of effects \citep[see][]{athey2022design,arkhangelsky2021double}. The previous display shows that those results do not extend to heteroscedastic and positively-serially-correlated non-binary treatments.

\paragraph{Standardizing the shocks can eliminate the negative weights.} Assume again that $V\left(Z_{s,t}\right)=\sigma^2_t$ and $\cov\left(Z_{s,1},Z_{s,2}\right)=\rho \sigma_1 \sigma_{2}$. Let $Z_{g,t}^{\text{sd}}=Z_{g,t}/\sigma_t$ denote the standardized Bartik instrument, and let
\begin{align*}
	\theta^{b,\text{sd}}&=\frac{\sum_{g=1}^G E\left(\Delta Y_{g} \left( \Delta Z_{g}^{\text{sd}}-E\left({\Delta Z}_{.}^{\text{sd}}\right)\right)\right) }{\sum_{g=1}^G E\left(\Delta D_{g} \left( \Delta Z_{g}^{\text{sd}}-E\left({\Delta Z}_{.}^{\text{sd}}\right)\right)\right)}
	\end{align*}
denote the estimand attached to a 2SLS regression of $\Delta Y_{g}$ on $\Delta D_{g}$ using $\Delta Z_{g}^{\text{sd}}$ as the instrument.
Under the assumptions of Theorem \ref{th_2sls_c_randomshocks}, one can show that
\begin{align*}
\theta^{b,\text{sd}}&=E\left(\sum_{g=1}^G \sum_{t=1}^2   \frac{\sigma_t\sum_{s=1}^S\beta_{s,g,t}Q^2_{s,g}}{E\left(\sum_{g'=1}^G \sum_{t'=1}^2 \sigma_{t'}\sum_{s=1}^S\beta_{s,g',t'}Q^2_{s,g'}\right)} \alpha_{g,t}\right),
\end{align*}
so unlike $\theta^{b}$, $\theta^{b,\text{sd}}$ is robust to heterogeneous treatment effects. Alternatively, combining the results in \cite{borusyak2020quasi} and \cite{adao2019shift} to those in \cite{angrist1998}, it follows that with randomly-assigned shocks, a 2SLS Bartik regression of $Y_{g,t}$ on $D_{g,t}$ using $Z_{g,t}$ as the instrument, with period fixed effects but no location fixed effects, estimates a weighted average of treatment effects, even if treatment effects vary across locations and over time.

\paragraph{Comparing Theorem \ref{th_2sls_c_randomshocks} to Proposition 3 in \cite{adao2019shift} and Proposition A.1 in \cite{borusyak2020quasi}.} Proposition 3 in \cite{adao2019shift} and Proposition A.1 in \cite{borusyak2020quasi} imply that with randomly-assigned shocks, cross-sectional 2SLS Bartik regressions are robust to heterogeneous effects.  To apply these results to the panel data case we consider here,
one can assume that the first-differenced variables $\Delta Y_{g}$, $\Delta D_{g}$, and $(\Delta Z_{s})_{s\in \{1,...,S\}}$ verify the assumptions underlying those results. This leads to the same decomposition as in \eqref{eq:resultAKM2}, under different assumptions. In particular, using this route, one can show that $\theta^{b}$ is robust to heterogeneous effects, even if effects are time varying, even if shocks are heteroscedastic and correlated, and even if $\cov\left(Z_{s,1},Z_{s,2}-Z_{s,1}\right)\ne 0$. However, as highlighted by \cite{borusyak2023}, the fundamental difference between our Theorem \ref{th_2sls_c_randomshocks} and
this direct application of Proposition 3 in \cite{adao2019shift} or Proposition A.1 in \cite{borusyak2020quasi} to first-differenced variables is that the former relies on a causal model in levels, while the latter relies on a causal model in first-difference. As shown in Lemma \ref{lem:causal_fd_levels}, having a causal model in first-difference with time-varying effects requires ruling out a causal model in levels, a strong requirement.

\paragraph{Randomly-assigned shocks, or randomly-assigned first-differenced shocks?} It is worth noting that Proposition 3 in \cite{adao2019shift} or Proposition A.1 in \cite{borusyak2020quasi}, if applied to $\Delta Y_{g}$, $\Delta D_{g}$, and $(\Delta Z_{s})_{s\in \{1,...,S\}}$, relies on Assumption \ref{as_randomly_assigned_shocks_fd} below. Assumption \ref{as_randomly_assigned_shocks_fd} is weaker than Assumption \ref{as_randomly_assigned_shocks}, as it requires that first-differenced shocks be as good as randomly assigned. Let $\mathcal{F}^{\text{fd}}=\left(\Delta Y_{g}(0),\Delta D_{g}(\bm{0}),\alpha_{g},(Q_{s,g},\beta_{s,g})_{s\in \{1,...,S\}}\right)_{g\in \{1,...,G\}}$.
 \begin{assumption}(Randomly-assigned first-differenced shocks) \label{as_randomly_assigned_shocks_fd}
\begin{enumerate}
\item For all $s$, $E\left(\Delta Z_{s}|\mathcal{F}^{\text{fd}}\right)=E\left(\Delta Z_{s}\right)$.
\item There exists a real number $\Delta \mu$ such that $E\left(\Delta Z_{s}\right)=\Delta\mu$ for all $s$.
\item The variables $\Delta Z_{s}$ are mutually independent across $s$, conditional on $\mathcal{F}^{\text{fd}}$.
\end{enumerate}
\end{assumption}

\paragraph{Testability of the randomly-assigned-shocks assumptions.} Finally, we highlight two testable implications of Assumption \ref{as_randomly_assigned_shocks_fd}, which to our knowledge had not been acknowledged so far. Assumption \ref{as_randomly_assigned_shocks} has similar testable implications, with shocks in levels. As Assumption \ref{as_randomly_assigned_shocks} is stronger than Assumption \ref{as_randomly_assigned_shocks_fd}, if one rejects Assumption \ref{as_randomly_assigned_shocks_fd} one can also reject Assumption \ref{as_randomly_assigned_shocks}. First, Point 2 of Assumption \ref{as_randomly_assigned_shocks_fd} implies that the expectation of $\Delta Z_{s}$ should not vary with sector-level characteristics, which can for instance be tested by regressing $\Delta Z_{s}$ on such characteristics. This test is similar to but different from that proposed by \cite{borusyak2020quasi}, who propose to regress each sector-level characteristic on $\Delta Z_{s}$. If that test is rejected for a sector-level covariate $X_s$, \cite{borusyak2020quasi} propose a remedy, which amounts to controlling for $\sum_{s=1}^SQ_{s,g}X_s$ in the Bartik regression. A limit of that strategy is that when shocks are correlated with some observables, shocks may also be correlated with some unobservables one cannot control for. The second testable implication we uncover is that Point 1 of Assumption \ref{as_randomly_assigned_shocks_fd} implies that $\Delta Z_{s}$ should be mean independent of the entire vector of shares $(Q_{s,g})_{g\in \{1,...,G\}}$, which implies
\begin{equation}\label{eq:test_implication_randomshocks1}
E\left(\Delta Z_{s}\middle|\frac{1}{G}\sum_{g=1}^G Q_{s,g}\right)=E\left( \Delta Z_{s}\right),
\end{equation}
an implication that can easily be tested, for instance by regressing first-differenced shocks on sectors' average share. If shocks are correlated with shares, the Bartik instrument can suffer from a standard endogeneity bias, even under constant treatment effects. For instance, if $\Delta Z_{s}$ is positively correlated with $\frac{1}{G}\sum_{g=1}^G Q_{s,g}$, $\Delta Z_{s}$ tends to be larger in sectors with a large average share, and locations with a larger-than-average share in sectors with a large average share will have a larger expectation of their first-differenced Bartik instrument than other locations. \cite{borusyak2020quasi} do not discuss a remedy for Bartik regressions with correlated shocks and shares. Proposing one such remedy goes beyond the scope of this paper, and may be intrinsically hard. First, as $\frac{1}{G}\sum_{g=1}^G Q_{s,g}$ is a function of the shares used to create locations' Bartik instruments, the aforementionned controlling strategy proposed by \cite{borusyak2020quasi} may not readily apply to that specific sector-level covariate. Even if that strategy does apply to that specific covariate, one could still be concerned that controlling for $\frac{1}{G}\sum_{g=1}^G Q_{s,g}$ is not enough: $\Delta Z_{s}$ may still be correlated with $(Q_{s,g})_{g\in \{1,...,G\}}$ conditional on $\frac{1}{G}\sum_{g=1}^G Q_{s,g}$.

\section{IV-CRC estimator}\label{sec_alternativeestimator}

In this section, we no longer assume that the instrument satisfies Definition \ref{def_bartik}. Our IV-CRC estimator is applicable whenever one has panel data and an instrument satisfying the assumptions below, irrespective of whether this instrument has a shift-share structure.

\paragraph{Group-level panel data set with at least three time periods.} In this section, we propose alternative estimators to FD 2SLS regressions. They build upon the correlated-random-coefficients (CRC) estimator proposed by \cite{chamberlain1992efficiency}. They can be used when the data has at least three periods.\footnote{With two periods, one may be able to follow a similar estimation strategy as that proposed in \cite{graham2012identification} and \cite{de2022continuous}.} For all $t \geq 2$ and any variable $R_{g,t}$, let $\Delta R_{g,t}=R_{g,t}-R_{g,t-1}$, and let $\bm{R}_{g}=( R_{g,1},...,R_{g,T})$ be a vector stacking the full time series of $R_{g,t}$. Our decompositions of FD 2SLS regressions extend to the multi-period case, as we show in Appendix \ref{sec_multipleperiods}.

\paragraph{Assumptions underlying our IV-CRC estimator.}
\begin{assumption}\label{as_additivsep_FS2S}
		For all $g \in \{1,...,G\}$ and $t\in \{1,...,T\}$, there exists real numbers $\lambda_t$ and random variables $\alpha_g$ such that $\alpha_{g,t}=\alpha_{g}+\lambda_t$.
\end{assumption}
Assumption \ref{as_additivsep_FS2S} allows for location-specific and time-varying effects, provided the treatment effects follow the same evolution over time in every location. Without loss of generality, we normalize $\lambda_1$ to $0$. Under Assumption \ref{as_additivsep_FS2S},
\begin{equation*}
\alpha_{ate}=E\left(\frac{1}{G}\sum_{g=1}^G\alpha_g\right)+\frac{1}{T}\sum_{t=1}^T\lambda_t
\end{equation*}
so identifying $\alpha_{ate,1}\equiv E\left(\frac{1}{G}\sum_{g=1}^G\alpha_g\right)$ and $(\lambda_2,...,\lambda_T)$ is sufficient to identify $\alpha_{ate}$. Assumption \ref{as_additivsep_FS2S} may be testable, if the data contains at least four time periods. Then, one can compute separately the IV-CRC estimator from periods one to three and from periods two to four, and verify if the average treatment effect follows the same evolution from period one to four across different subgroups of locations, though it is unclear how such subgroups should be formed. Formalizing this testing idea is left for future work.
\begin{assumption}\label{as_GMM_commontrend_y}
	For all $t \in \{2,...,T\}$, there are real numbers $\mu_t$ such that $\forall g\in \{1,...,G\}$, $E(\Delta Y_{g,t} (0)| \bm{Z}_{g})=\mu_t$.
\end{assumption}
Assumption \ref{as_GMM_commontrend_y} requires that locations' outcome evolutions without treatment be mean-independent of the full sequence of their instruments. Like the first point of Assumption \ref{as_commontrend_y}, it may be interpreted as a parallel trends assumption. However, Assumption \ref{as_GMM_commontrend_y} is stronger than that condition: it requires that $\Delta Y_{g,t}(0)$ be mean independent from $(Z_{g,1},...,Z_{g,T})$ rather than uncorrelated with $\Delta Z_{g,t}$. If the data contains a period $t_0\in \{2,...,T\}$ such that $D_{g,t_0}=D_{g,t_0-1}=0$ for all $g$, then $\Delta Y_{g,t_0} (0)$ is observed, and Assumption \ref{as_GMM_commontrend_y} has the following testable implication:
\begin{align}
E(\Delta Y_{g,t_0}|\bm{Z}_{g})=E(\Delta Y_{g,t_0}).\label{eq:testable_implication1}
\end{align}
To test \eqref{eq:testable_implication1}, one can for instance regress $\Delta Y_{g,t_0}$ on $Z_{g,t'}$ for any $t' \ne t_0$. One could also regress $\Delta Y_{g,t_0}$ on a polynomial in $(Z_{g,1},...,Z_{g,t_0-2},Z_{g,t_0+1},...,Z_{g,T})$.
\begin{assumption}\label{as_GMM_ignorable_treatment}
	For all $t$ and $g$, $E(\alpha_g|D_{g,t},\bm{Z}_{g})=E(\alpha_g|\bm{Z}_{g})$.
\end{assumption}
Assumption \ref{as_GMM_ignorable_treatment} requires that locations' treatment effects be independent of $D_{g,t}$, conditional on $\bm{Z}_{g}$: locations with the same vector of instruments but different values of $D_{g,t}$ should not have systematically different treatment effects. $\alpha_g$ is unobserved, so Assumption \ref{as_GMM_ignorable_treatment} is untestable. Still, if one observes covariates $X_g$ that are likely to be correlated with locations' treatment effects, one can suggestively test Assumption \ref{as_GMM_ignorable_treatment}, by regressing $X_g$ on $(D_{g,t},\bm{Z}_{g})$.

\paragraph{Identification result.}
Let $\tilde{D}_{g,t}=E(D_{g,t}|\bm{Z}_{g}).$ For all $t\geq 2$, let $\mu_{1:t}=\sum_{k=2}^t \mu_k$. Let $\bm{\theta}=(\mu_{1:2},\lambda_2,\mu_{1:3},\lambda_3,...,\mu_{1:T},\lambda_T)'$, let $\bm{0}_{k}$ denote a vector of $k$ zeros,
let
$$\mathcal{P}_g = \begin{pmatrix} \bm{0}_{2T-2}\\ 1, \tilde{D}_{g,2},\bm{0}_{2T-4}\\ \bm{0}_2, 1, \tilde{D}_{g,3},\bm{0}_{2T-6}\\ \vdots  \\\bm{0}_{2T-4}, 1, \tilde{D}_{g,T}\end{pmatrix}\hspace{0.25cm}\text{and }\mathcal{X}_{g} = \begin{pmatrix} 1, \tilde{D}_{g,1}\\ 1, \tilde{D}_{g,2}\\ \vdots  \\1, \tilde{D}_{g,T}\end{pmatrix}.$$
For any $T\times K$ matrix $A$, let $A^+$ be its Moore-Penrose inverse, and let $M(A)=\bm{I}_T-AA^+$ be the orthogonal projector on the kernel of $A$. For any $K\times 1$ vector $x$, $(x)_k$ is its $k$th coordinate.
\begin{theorem}\label{th_chamberlain}
Suppose that Assumptions \ref{as_linear_ss} and \ref{as_additivsep_FS2S}-\ref{as_GMM_ignorable_treatment} hold, $E\left(\frac{1}{G}\sum_{g=1}^G \mathcal{P}_g'M(\mathcal{X}_{g})\mathcal{P}_g\right)$ is invertible, and with probability 1 $\mathcal{X}_{g}'\mathcal{X}_{g}$ is invertible for every $g\in \{1,...,G\}$. Then:
\begin{align}
\bm{\theta}=&E\left(\frac{1}{G}\sum_{g=1}^G \mathcal{P}_g'M(\mathcal{X}_{g})\mathcal{P}_g\right)^{-1}E\left(\frac{1}{G}\sum_{g=1}^G \mathcal{P}_g'M(\mathcal{X}_{g})\bm{ Y}_{g}\right),\label{eq:thetaY}\\
\alpha_{ate,1}=&\left(E\left(\frac{1}{G}\sum_{g=1}^G\left(\mathcal{X}_{g}'\mathcal{X}_{g}\right)^{-1}\mathcal{X}_{g}'\left(\bm{ Y}_{g}-\mathcal{P}_g\bm{\theta}\right)\right)\right)_2\label{eq:alphag}.
\end{align}
\end{theorem}
\paragraph{Estimation.} We estimate $\alpha_{ate}$ under a functional-form assumption on $E(D_{g,t}|\bm{Z}_{g})$.
\begin{assumption}\label{as_GMM_functional_form}
	There exists an integer $K$ such that for all $t\geq 2$, there is a polynomial of order $K$ and of $T$ variables $P_{K,t}$ such that for all $g$, $E(D_{g,t}|\bm{Z}_{g})=P_{K,t}(\bm{Z}_{g})$.
\end{assumption}
Polynomials are well suited to a large class of applications, but when they are not one can of course assume a different functional form. Under Assumption \ref{as_GMM_functional_form}, one may estimate $\alpha_{ate}$ as follows. First, one regresses $D_{g,t}$ on a polynomial of order $K$ in $\bm{Z}_{g}$, separately for every $t\geq 2$. Then, letting $\widehat{\tilde{D}}_{g,t}$ denote the prediction from that estimation, one lets
$$\widehat{\mathcal{P}}_g = \begin{pmatrix} \bm{0}_{2T-2}\\ 1, \widehat{\tilde{D}}_{g,2},\bm{0}_{2T-4}\\ \bm{0}_2, 1, \widehat{\tilde{D}}_{g,3},\bm{0}_{2T-6}\\ \vdots  \\\bm{0}_{2T-4}, 1, \widehat{\tilde{D}}_{g,T}\end{pmatrix}\hspace{0.25cm}\text{and }\widehat{\mathcal{X}}_{g} = \begin{pmatrix} 1, \widehat{\tilde{D}}_{g,1}\\ 1, \widehat{\tilde{D}}_{g,2}\\ \vdots  \\1, \widehat{\tilde{D}}_{g,T}\end{pmatrix},$$
and
\begin{align*}
\widehat{\bm{\theta}}=&\left(\frac{1}{G}\sum_{g=1}^G \widehat{\mathcal{P}}_g'M(\widehat{\mathcal{X}}_{g})\widehat{\mathcal{P}}_g\right)^{-1}\left(\frac{1}{G}\sum_{g=1}^G \widehat{\mathcal{P}}_g'M(\widehat{\mathcal{X}}_{g})\bm{ Y}_{g}\right),\\
\widehat{\alpha}_{ate,1}=&\left(\frac{1}{G}\sum_{g=1}^G\left(\widehat{\mathcal{X}}_{g}'\widehat{\mathcal{X}}_{g}\right)^{-1}\widehat{\mathcal{X}}_{g}'\left(\bm{ Y}_{g}-\widehat{\mathcal{P}}_g\widehat{\bm{\theta}}\right)\right)_2,\\
\widehat{\alpha}_{ate}=&\widehat{\alpha}_{ate,1}+\frac{1}{T}\sum_{t=1}^T\widehat{\lambda}_t.
\end{align*}
Estimating $\alpha_{ate}$ without a functional-form assumption on $E( D_{g,t}| \bm{Z}_{g})$ is feasible, using a non-parametric estimator of $E( D_{g,t}| \bm{Z}_{g})$. We leave this extension for future work.

\paragraph{Intuition.} Our estimator may be seen as an IV-version of Chamberlain's CRC estimator. In a first step, one uses the vector of instruments $\bm{ Z}_{g}$ to predict the treatment $D_{g,t}$. Then, one computes the CRC estimator with the predicted treatment in lieu of the endogenous treatment. To simplify the presentation of the identification argument, we momentarily assume that $T=3$, and that treatment effects are location-specific but time invariant: $\alpha_{g,t}=\alpha_g$. Then,
\begin{align}
E(\Delta Y_{g,t}|\bm{ Z}_{g})=&E(\Delta Y_{g,t}(0)|\bm{ Z}_{g})+E(\alpha_g \Delta D_{g,t}|\bm{ Z}_{g})\nonumber\\
=&\mu_t+E(\alpha_g |\bm{ Z}_{g})\Delta \tilde{D}_{g,t},\label{eq:id_CRC}
\end{align}
where the second equality follows from Assumptions \ref{as_GMM_commontrend_y} and \ref{as_GMM_ignorable_treatment}.
Then, subtracting \eqref{eq:id_CRC} at $t=3$ multiplied by $\Delta\tilde{D}_{g,2}\Delta\tilde{D}_{g,3}$ from \eqref{eq:id_CRC} at $t=2$ multiplied by $\Delta\tilde{D}^2_{g,3}$ yields
\begin{align}
\Delta\tilde{D}^2_{g,3}E(\Delta Y_{g,2}|\Delta \bm{ Z}_{g})-\Delta\tilde{D}_{g,2}\Delta\tilde{D}_{g,3}E(\Delta Y_{g,3}|\Delta \bm{ Z}_{g})=&\Delta\tilde{D}^2_{g,3}\mu_2-\Delta\tilde{D}_{g,2}\Delta\tilde{D}_{g,3}\mu_3,\label{eq:id_CRC2}
\end{align}
an equation that does not depend on the treatment effect.
Similarly, subtracting \eqref{eq:id_CRC} at $t=2$ multiplied by $\Delta\tilde{D}_{g,2}\Delta\tilde{D}_{g,3}$ from \eqref{eq:id_CRC} at $t=3$ multiplied by $\Delta\tilde{D}^2_{g,2}$ yields
\begin{align}
\Delta\tilde{D}^2_{g,2}E(\Delta Y_{g,3}|\Delta \bm{ Z}_{g})-\Delta\tilde{D}_{g,2}\Delta\tilde{D}_{g,3}E(\Delta Y_{g,2}|\Delta \bm{ Z}_{g})=&\Delta\tilde{D}^2_{g,2}\mu_3-\Delta\tilde{D}_{g,2}\Delta\tilde{D}_{g,3}\mu_2,\label{eq:id_CRC3}
\end{align}
an equation that also does not depend on the treatment effect. \eqref{eq:id_CRC2} and \eqref{eq:id_CRC3} give a system of conditional moment equalities with two unknowns,
$\mu_2$ and $\mu_3$, so $\mu_2$ and $\mu_3$ are identified. Then, it follows from \eqref{eq:id_CRC} that $E(\alpha_g |\Delta \bm{ Z}_{g})$ is identified.\footnote{Applying results in \cite{chamberlain1992efficiency}, one can derive the optimal estimator of $(\mu_2,\mu_3)$ attached to this system of conditional moment equalities. An issue, however, is that Chamberlain's optimality results do not apply to the estimators of $\alpha_{ate,1}$ and $\lambda_t$, the building blocks of our target parameter. Moreover, the computation of the optimal estimator requires a non-parametric first-stage estimation. To our knowledge, no data-driven method has been proposed to choose the tuning parameters involved in this first stage. Accordingly, we prefer to stick with estimators of $(\mu_2,\mu_3)$ attached to unconditional moment equalities.}

\paragraph{Inference.} We suggest a method to draw inference on the ATE under Assumption \ref{as_GMM_indepgroups}.
\begin{assumption}\label{as_GMM_indepgroups}
	$\left(\bm{Z}_{g},\bm{D}_{g},\bm{Y}_{g}\right)_{g}$ is iid.
\end{assumption}
Assumption \ref{as_GMM_indepgroups} requires that the instruments, treatments, and outcomes be iid across locations, but it allows for serial correlation within locations. Under Assumption \ref{as_GMM_indepgroups}, to estimate the standard error of $\widehat{\alpha}_{ate}$, we propose to bootstrap the whole estimation procedure, clustering the bootstrap at the location level.
The heuristic identification argument above shows that the estimand identifying the ATE involves third moments of $(\Delta Y_{g,2},\Delta Y_{g,3},\Delta\tilde{D}_{g,2},\Delta\tilde{D}_{g,3})$, while the FD 2SLS estimand only involves first and second moments. This may explain why when we revisit ADH, the variance of the IV-CRC estimator is substantially larger than that of the FD 2SLS estimator. Noteworthy, using the IV-CRC estimator instead of the FD 2SLS one does not always lead to precision losses as large as those we find in ADH: when we revisit the canonical Bartik design, the variance of the IV-CRC estimator is slightly larger than that of the FD 2SLS estimator, but the difference is much lower than in ADH (see Web Appendix Tables \ref{table:canonical_Bartikreg} \& \ref{table:BTK_IV-CRC}).

\paragraph{Inference in Bartik designs.}
In Bartik designs, Assumption \ref{as_GMM_indepgroups} can only hold conditional on the shocks. Thus, it is compatible with the shares approach, not with the shocks one.\footnote{If the treatment and outcome are also influenced by unobserved sector-level shocks, as hypothesized in \cite{borusyak2020quasi} and \cite{adao2019shift}, those shocks need to be conditioned upon for Assumption \ref{as_GMM_indepgroups} to be plausible.} Accordingly, the bootstrapped standard error we propose does not account for the variance arising from the shocks. Accounting for it would require extending the approach in \cite{adao2019shift} to the estimators in Theorem \ref{th_chamberlain}. This extension is left for future work.

\paragraph{FD 2SLS regressions are still not robust to heterogeneous treatment effects under the assumptions underlying our IV-CRC estimator.} Under Assumptions \ref{as_linear_ss} and \ref{as_additivsep_FS2S}-\ref{as_GMM_ignorable_treatment}, if one further assumes constant effects over time ($\lambda_t=0$), it follows from Point 2 of Theorem \ref{th_2sls_c} that when $T=2$, $\theta^b$ identifies a weighted sum of the conditional effects $E(\alpha_g|\bm{Z}_g)$, potentially with some negative weights, proportional to $E(\Delta D_g|\bm{Z}_g)(\Delta Z_g-E(\Delta Z_.))$ (a similar result holds for $T>2$). Thus, $\theta^b$ may not be robust to heterogeneous effects across locations, under stronger assumptions than those under which our IV-CRC estimand identifies the ATE.

\paragraph{Estimator with control variables.}
Let $X_g$ be a $K\times 1$ vector of time-invariant location-level control variables, with $k$th coordinate $X_{k,g}$. An IV-CRC estimator controlling for $X_g$ can be obtained, replacing Assumptions \ref{as_GMM_commontrend_y} and \ref{as_GMM_ignorable_treatment} by the following conditions:
\begin{assumption}\label{as_GMM_commontrend_y_X}
	For all $t \in \{2,...,T\}$, there is a  real number $\mu_t$ and a $K\times 1$ vector  $\mu_{X,t}$ such that $\forall g\in \{1,...,G\}$, $E(\Delta Y_{g,t}(0)| \bm{Z}_{g},X_g)=E(\Delta Y_{g,t}(0)|X_g)=\mu_t+X'_g\mu_{X}$.
\end{assumption}
\begin{assumption}\label{as_GMM_ignorable_treatment_X}
	For all $t \in \{2,...,T\}$ and $g$, $E(\alpha_g|D_{g,t},\bm{Z}_{g},X_g)=E(\alpha_g|\bm{Z}_{g},X_g)$.
\end{assumption}
Assumption \ref{as_GMM_commontrend_y_X} may be more plausible than Assumption \ref{as_GMM_commontrend_y}: it requires that locations' outcome evolutions without treatment be mean-independent of their instruments conditional on $X_g$, rather than unconditionally. Then, the second equality requires that $E(\Delta Y_{g,t}(0)|X_g)$ be linear in $X_g$. Assumption \ref{as_GMM_ignorable_treatment_X} may also be more plausible than Assumption \ref{as_GMM_ignorable_treatment}. For all $t\geq 2$, let $\mu_{X,1:t}=\sum_{k=2}^t \mu_{X,k}$. Redefining $\tilde{D}_{g,t}\equiv E(D_{g,t}|\bm{Z}_{g},X_g)$,
$\bm{\theta}\equiv(\mu_{1:2},\lambda_2,\mu_{1:3},\lambda_3,...,\mu_{1:T},\lambda_T,\mu_{X,1:2},...,\mu_{X,1:T})',$
$$\mathcal{P}_g \equiv  \begin{pmatrix} \bm{0}_{2T-2+(T-1)K}\\ 1, \tilde{D}_{g,2},\bm{0}_{2T-4},X_{1,g},...,X_{K,g},\bm{0}_{(T-2)K}\\ \bm{0}_2, 1, \tilde{D}_{g,3},\bm{0}_{2T-6+K},X_{1,g},...,X_{K,g},\bm{0}_{(T-3)K}\\ \vdots  \\\bm{0}_{2T-4}, 1, \tilde{D}_{g,T},\bm{0}_{(T-2)K},X_{1,g},...,X_{K,g}\end{pmatrix},$$
one can show that \eqref{eq:thetaY} and \eqref{eq:alphag} still hold under Assumptions \ref{as_linear_ss}, \ref{as_additivsep_FS2S},  \ref{as_GMM_commontrend_y_X}, and \ref{as_GMM_ignorable_treatment_X}.

\paragraph{Estimator assuming constant effects over time.} Similarly, it is easy to obtain an IV-CRC estimator assuming that treatment effects are constant over time. Then, \eqref{eq:thetaY} and \eqref{eq:alphag} still hold, after redefining $\bm{\theta}\equiv(\mu_{1:2},0,\mu_{1:3},0,...,\mu_{1:T},0)',$ and
$$\mathcal{P}_g \equiv \begin{pmatrix} \bm{0}_{2T-2}\\ 1, 0,\bm{0}_{2T-4}\\ \bm{0}_2, 1, 0,\bm{0}_{2T-6}\\ \vdots  \\\bm{0}_{2T-4}, 1, 0\end{pmatrix}.$$

\paragraph{Testing for heterogeneous treatment effects.} In the proof of Theorem \ref{th_chamberlain}, we show that
$$E(\alpha_{g})=\left(E\left(\left(\mathcal{X}_{g}'\mathcal{X}_{g}\right)^{-1}\mathcal{X}_{g}'\left(\bm{ Y}_{g}-\mathcal{P}_g\bm{\theta}\right)\right)\right)_2,$$
a result stronger than that in \eqref{eq:alphag}. Then one may use
\begin{equation}\label{eq:alphaghat}
\widehat{\alpha}_{g}\equiv \left(\left(\widehat{\mathcal{X}}_{g}'\widehat{\mathcal{X}}_{g}\right)^{-1}\widehat{\mathcal{X}}_{g}'\left(\bm{ Y}_{g}-\widehat{\mathcal{P}}_g\widehat{\bm{\theta}}\right)\right)_2
\end{equation}
to estimate $E(\alpha_{g})$, and $\sum_t (\widehat{\alpha}_g+\widehat{\lambda}_t)$ to estimate an effect specific to CZ $g$, on average across all time periods. With a fixed number of time periods $T$, those estimators
are not consistent, and naively using them to estimate the distribution of treatment effects across locations would be misleading: one would first need to deconvolute them. Proposing a deconvolution technique goes beyond the scope of this paper. Another possibility to test for heterogeneous effects is to regress $\sum_t (\widehat{\alpha}_g+\widehat{\lambda}_t)$ on location-level covariates, and assess whether the covariates significantly predict those estimates effects \citep[see][who made a similar proposal before this paper]{muris2022estimating}. Inference still needs to account for the fact the $\widehat{\alpha}_{g}$s are estimated, which may be achieved by bootstrapping the estimation procedure.

\section{Empirical application: China shock}\label{sec_appliChina}

\subsection{Treatment, outcome, and instrument in ADH}

\paragraph{Structural equation guiding the treatment definition in ADH.}
In their Web Appendix, ADH consider a small open economy model, from which they derive a structural equation, Equation (2) in their paper, that guides their treatment definition. Their Equation (2) is a first-order Taylor approximation where the difference between two counterfactual levels of the logarithm of total employment in traded goods in CZ $g$ at $t$\footnote{The model in the Web Appendix of ADH has only one period, but with several periods one can re-derive its equilibrium equations at each period.} under two different vectors of log Chinese export-supply capabilities $(\ln(esc^a_{s,t}))_{s\in \{1,...,S\}}$ and $(\ln(esc^b_{s,t}))_{s\in \{1,...,S\}}$ is expressed as a linear function of
a weighted average of $(\ln(esc^a_{s,t})-\ln(esc^b_{s,t}))/E_{s,t}$, where $E_{s,t}$ is the number of US workers in sector $s$ at $t$. Specifically,
\begin{equation*}
\ln\left(Y_{g,t}\left((\ln(esc^a_{s,t}))_{s\in \{1,...,S\}}\right)\right)-\ln\left(Y_{g,t}\left((\ln(esc^b_{s,t}))_{s\in \{1,...,S\}}\right)\right)\approx \alpha_{g,t} \sum_{s=1}^S Q_{s,g,t}\frac{\ln(esc^a_{s,t})-\ln(esc^b_{s,t})}{E_{s,t}},
\end{equation*}
where $Q_{s,g,t}$ is the share of sector $s$ in the total employment in traded goods of CZ $g$ at $t$. Letting $\ln(esc^b_{s,t})=0$ for all $s$, and letting $$c_{g,t}=\sum_{s=1}^S Q_{s,g,t}\frac{\ln(esc_{s,t})}{E_{s,t}}$$
denote $(g,t)$'s exposure to China's export supply capability,
the previous display implies
\begin{equation*}
\ln\left(Y_{g,t}\left(c_{g,t}\right)\right)\approx \ln\left(Y_{g,t}\left(0\right)\right)+\alpha_{g,t}c_{g,t},
\end{equation*}
a linear causal model relating the logarithm of counterfactual total employment in traded goods in CZ $g$ at $t$ to $c_{g,t}$. In the previous display, $\alpha_{g,t}$, the effect of $c_{g,t}$, is allowed to vary across $g$ and $t$, while that effect is constant in Equation (2) in ADH. Their model can deliver a linear causal model with heterogeneous treatment effects, for instance if one allows
the trade imbalance in total expenditure, $\rho$ in their notation, to vary across CZs and over time, as is likely the case in reality (see their Equation (1)). Importantly, Equation (2) in ADH is a model for $\ln\left(Y_{g,t}\left(c_{g,t}\right)\right)$ and not for the first-difference of that variable.

\paragraph{Treatment definition.}
Obviously, sectoral Chinese export-supply capabilities are unobserved. Let $M^{US}_{s,t}$ denote US imports from China in sector $s$ at $t$. ADH define their treatment as
\begin{equation}\label{eq:treat_obs_Autor}
	D_{g,t}\equiv\sum_{s=1}^S Q_{s,g,t}\frac{M^{US}_{s,t}}{E_{s,t}}.
\end{equation}
This treatment definition follows that of $c_{g,t}$ above, except that the unobserved log-export-supply capabilities are replaced by the observed sectoral imports from China.\footnote{ADH do not take the log of exports when they define their treatment, which might have been natural in view of $c_{g,t}$'s definition, probably because exports can be equal to zero.}

\paragraph{Outcome definition.} The main outcome variable $Y_{g,t}$ in ADH is the manufacturing employment share of the working-age population in CZ $g$ at $t$, hereafter referred to as the ``manufacturing employment share''.\footnote{Again, that variable is not in logs, probably to be consistent with the fact that exports are not in logs.}

\paragraph{Instrument definition.} $M^{US}_{s,t}$ is determined by China's export supply capability, but also by US demand in $s$ at $t$. This may create a correlation between $D_{g,t}$ and other determinants of $Y_{g,t}$ than China's exports. Accordingly, ADH define the following instrument:
\begin{equation}\label{eq:instrum_obs_Autor}
	Z_{g,t}\equiv\sum_{s=1}^S Q_{s,g,t} \frac{M^{OC}_{s,t}}{E_{s,t}},
\end{equation}
where $M^{OC}_{s,t}$ denotes China's exports in sector $s$ at $t$ to eight high-income countries similar to the US, hereafter referred to as other countries.

\medskip
In Appendix \ref{sec_theorymetrics}, we give an economic interpretation of our econometric assumptions in ADH, under a gravity-based decomposition of trade flows.

\subsection{Data and variables' definitions}

\paragraph{Data.} We use the replication dataset of ADH on the AEA website. In their main analysis, they use a CZ-level panel data set, with 722 CZs and 3 periods (1990, 2000, and 2007). This data set does not contain the shock and share variables. We obtained those variables from the replication dataset of \cite{borusyak2020quasi}.

\paragraph{Time-invariant shares and sectoral employments.} While variables are in levels in their theoretical model, in their statistical analysis ADH define variables directly in first-differences (see their Equations (3) and (4)). Their first-differenced treatment is\footnote{In this paper, $\Delta D_{g,t}=D_{g,t}-D_{g,t-1}$, while in Equation (3) in ADH $\Delta D_{g,t}=D_{g,t+1}-D_{g,t}$. Therefore, our $\Delta D_{g,t}$ coincides with $\Delta D_{g,t-1}$ in Equation (3) in ADH. The same applies to $\Delta Z_{g,t}$ defined below: it coincides with $\Delta Z_{g,t-1}$ in Equation (4) in ADH.}
\begin{equation}\label{eq:deltad_def_ADH}
	\Delta D_{g,t}=\sum_{s=1}^S Q_{s,g,t-1} \times \frac{\Delta M^{US}_{s,t}}{E_{s,t-1}},
\end{equation}
where $Q_{s,g,t-1}$ is the employment share of $s$ in $g$ at $t-1$. This first-differenced treatment does not coincide with the first-difference of \eqref{eq:treat_obs_Autor}. Similarly, the first-differenced instrument is defined as
\begin{equation*}
	\Delta Z_{g,t}=\sum_{s=1}^S Q_{s,g,t-2} \times \frac{\Delta M^{OC}_{s,t}}{E_{s,t-2}}.
\end{equation*}
We need to define a treatment in levels, because the weights in our decomposition of $\theta^b$ in Point 1 of Theorem \ref{th_2sls_c} depend on it. Therefore, we use time-invariant shares and sectoral employments, to construct consistent levels and first-differences of the treatment and instrument. Shares and sectoral employment are set at their 1980 value for the instrument, and at their 1990 value for the treatment, to reflect the fact ADH use lagged shares and sectoral employments for the instrument.

\paragraph{Extrapolated decennial panel.}
The trade data used by ADH is available in 1991, 2000, and 2007. To construct first-differenced variables over a comparable time span, ADH multiply their 1991-2000 first-differenced variables by 10/9, and their 2000-2007 first-differenced variables by 10/7. We adopt the same strategy to extrapolate variables in levels. Specifically, for $\text{dest}\in \{US,OC\}$, we let $M^{dest}_{s,1990}= M^{\text{dest}}_{s,2000}-10/9( M^{\text{dest}}_{s,2000}-\Delta M^{\text{dest}}_{s,1991})$, and $M^{\text{dest}}_{s,2010}= M^{\text{dest}}_{s,2000}+10/7(M^{\text{dest}}_{s,2007}- M^{\text{dest}}_{s,2000})$. We adopt a similar strategy to construct CZs extrapolated employment level in 2010, as the last employment measurement in ADH uses the 2006, 2007, and 2008 American Community Survey (ACS).

\paragraph{Comparing our variables with those in ADH.}
Our 1990-to-2000 first-differenced instrument, treatment, and outcome take exactly the same values as in the original ADH dataset. Our 2000-to-2010 first-differenced outcome also takes exactly the same values as in the original data. On the other hand, our 2000-to-2010 first-differenced treatment and instrument differ slightly from those in the original ADH dataset, as we use fixed shares and sectoral employments while ADH use time-varying ones. The correlation between our and ADH's 2000-to-2010 first-differenced treatment is 0.749 (p-value$<$0.001), and the correlation between our and ADH's 2000-to-2010 first-differenced instrument is 0.820 (p-value$<$0.001). We will show below that our main results are not driven by the fact we slightly change ADH variables' definitions.

\subsection{Tests of the identifying assumptions}\label{subsec_testChina}

\subsubsection{The randomly-assigned shocks assumption is rejected}

Below, we test Points 1 and 2 of Assumption \ref{as_randomly_assigned_shocks_fd}: as they are weaker than Points 1 and 2 of Assumption \ref{as_randomly_assigned_shocks}, if we reject the former we can also reject the latter.

\paragraph{Shocks' first-differences are correlated to sectors' average shares.} Point 1 of Assumption \ref{as_randomly_assigned_shocks_fd} implies that $E\left(\Delta Z_{s,t}|\frac{1}{G}\sum_{g=1}^G Q_{s,g}\right)=E\left(\Delta Z_{s,t}\right)$: first-differenced shocks should be mean independent of the average share of sector $s$ across locations. We test this by regressing $\Delta Z_{s,t}$ on $\frac{1}{G}\sum_{g=1}^G Q_{s,g}$ in Panel A of Table \ref{table:ADH_testrandomshocks}, for $t=2000$ in Column (1) and for $t=2010$ in Column (2). We follow Table 3 Panel A in \cite{borusyak2020quasi}, and cluster standard errors at the level of three-digit SIC codes, but results are very similar when one uses robust standard errors. We reject the null, with t-stats equal to -1.96 and -3.93 in Columns (1) and (2): large shocks are more likely to arise in sectors with a lower average share. Results are similar if we use the first-differenced shocks and shares defined by \cite{borusyak2020quasi}, rather than our variables.

\paragraph{Shocks' first-differences are correlated to sectors' characteristics.} Point 2 of Assumption \ref{as_randomly_assigned_shocks_fd} implies that the expectation of shocks' first-differences should not vary with sector-level characteristics. We test this by regressing shocks' first-differences on such characteristics. We use the five sector characteristics in \cite{acemoglu2016import} that are in the replication dataset of \cite{borusyak2020quasi}.
Panel B of Table \ref{table:ADH_testrandomshocks} shows regressions of shocks' first-differences from 1990 to 2000 and from 2000 to 2010 on these characteristics. We follow Table 3 Panel A in \cite{borusyak2020quasi} and weight the regressions by sectors' average shares, but the results are very similar when the regressions are not weighted. We find that large shocks' first-differences tend to appear in sectors with low wages and more computer and high-tech investment. We can reject the hypothesis that shocks' first-differences are not correlated with any sectoral characteristic (p-value$<$0.001 in Column (1), p-value$=$0.038 in Column (2)). Results are similar if we use the first-differenced shocks defined by \cite{borusyak2020quasi}, rather than our variables.

\paragraph{Conditionally randomly assigned shocks?} Shocks could be as-good-as randomly assigned conditional on sectoral characteristics. If that were true, shocks first-differences should be mean independent of sectors' average shares conditional on those characteristics. We can test this, by adding sectors' average shares to the regressions shown in Panel B of Table \ref{table:ADH_testrandomshocks} (the regressions are no longer weighted by sectors' average shares). The coefficients on sectors' average share are highly significant (p-values$=$0.015 for $t=2000$, $=$0.004 for $t=2010$).

\paragraph{Comparison with the test of Assumption \ref{as_randomly_assigned_shocks_fd}  in \cite{borusyak2020quasi}.}
Our test of Assumption \ref{as_randomly_assigned_shocks_fd} in Panel B of Table \ref{table:ADH_testrandomshocks} is inspired from, and related to, that in Table 3 Panel A in \cite{borusyak2020quasi}. Regressing each sectoral characteristic on the shocks, they find no significant correlation between characteristics and shocks. As explained above, the difference between our and their results does not come from the differences in our variables' definitions. Reverting the dependent and the independent variables in their Table 3 Panel A would leave their t-stats unchanged, so the difference between our and their test is that they regress the shocks on each characteristic individually, while we regress the shocks on all the characteristics. It follows from standard OLS formulas that the null in our test is stronger than the null in their test: if the coefficients of all characteristics are equal to zero in our long regression, then the coefficients of all characteristics are equal to zero in their short regressions. The fact that we test a stronger implication of Assumption \ref{as_randomly_assigned_shocks} may explain why our test is rejected while theirs is not, though testing a stronger null does not always imply a larger finite-sample power.

\paragraph{Can we consider that shocks are as good as randomly assigned in ADH?} \cite{borusyak2023} argue that despite our results in Panel B of Table \ref{table:ADH_testrandomshocks}, as-good-as-random shock assignment can still be a reasonable assumption in ADH. To support their argument, they note that in their Table 4 Column (6), \cite{borusyak2020quasi} still find a significantly negative effect of imports from China on US employment, even controlling for the five sector-level covariates in Panel B of Table \ref{table:ADH_testrandomshocks} (though, interestingly, their point estimate is about twice smaller than that in ADH). In our opinion, there are two limits with their argument. First, in observational studies relying on the assumption that an instrument is randomly assigned, we believe that the primary goal of balancing checks is not to select the covariates that need to be controlled for so that the instrument is conditionally randomly assigned. Rather, we believe that balancing checks ought to confirm that the instrument is not correlated with some observables, so that one can be reasonably confident that the instrument is not correlated with some unobservables. In view of Panel B of our Table \ref{table:ADH_testrandomshocks}, we believe there is a legitimate concern that the point estimate of Table 4 Column (6) of \cite{borusyak2020quasi} may still be biased, because the instrument may still be correlated with some unobservables, even conditional on the observables controlled for in this specification. Second, and more importantly, \cite{borusyak2023} do not comment on the results in Panel A of our Table \ref{table:ADH_testrandomshocks}. However, the correlation between shocks and shares, that remains significant even conditional on all the covariates in Panel B of our Table \ref{table:ADH_testrandomshocks}, is further evidence that the random shocks assumption is violated in ADH. Moreover, \cite{borusyak2020quasi} do not propose a remedy for Bartik designs with correlated shares and shocks. As mentioned earlier, proposing one such remedy may be intrinsically hard. Therefore, in what follows we consider that the random-shocks assumption is rejected in ADH.

\begin{table}[H]
	\begin{center}
		\caption{\normalsize Testing the random first-differenced shocks assumption}
		\begin{tabular}{lcc}
			\hline
			& (1)       & (2)     \\
			Variables         & $\Delta Z_{s,t}$: 1990-2000    & $\Delta Z_{s,t}$: 2000-2010        \\
			\hline
			\textit{Panel A: Shocks uncorrelated to sectors' average share?} & & \\
			Sector's average share & -567.488 & -1,765.791 \\
			& (289.280) & (448.911) \\
			\hline
			\textit{Panel B: Shocks uncorrelated to sectors' characteristics?} & & \\
			Production workers' share of employment$_{1991}$   & 3.447 & 9.481    \\
			& (5.155)   & (18.260) \\
			Ratio of capital to value-added$_{1991}$     & -0.357  & 1.188 \\
			& (0.908)   & (2.222)  \\
			Log real wage (2007 USD)$_{1991}$ & -8.328 & -3.815  \\
			& (2.092)   & (6.117)  \\
			Computer investment as share of total investment$_{1990}$      & 0.173     & 1.058   \\
			& (0.113)   & (0.457)  \\
			High-tech equipment as share of total investment$_{1990}$   & 0.206   & 0.685  \\
			& (0.131)   & (0.370)  \\
			F-test P-value     &    0.0000       &  0.0383       \\ \hline
			Observations       & 397       & 397      \\
			\hline
		\end{tabular}
		\label{table:ADH_testrandomshocks}
	\end{center}
\footnotesize{Notes: The dependent variable in Column (1) (resp. (2)) is the change in per-worker imports from China to other high-income countries from 1990 to 2000 (resp. from 2000 to 2010). In Panel A, the independent variable is sectors' average shares across commuting zones. In Panel B, the independent variables are five sector characteristics obtained from \cite{acemoglu2016import}: sectors' share of production workers in employment in 1991, sectors' ratios of capital to value-added in 1991, sectors' log real wages in 1991, sectors' share of investment devoted to computers in 1990, and sectors' share of high-tech equipment in total investment in 1990. Standard errors clustered at the level of three-digit SIC codes are shown in parentheses. The regressions in Panel B are weighted by sectors' average shares. The F-test p-value in Panel B is the p-value of the test that the coefficients on all sector characteristics are equal to 0.}
\end{table}

\subsubsection{Assumption \ref{as_GMM_commontrend_y} is rejected, Assumption \ref{as_GMM_commontrend_y_X} is not.}

\paragraph{Interpretation of the placebo tests in ADH.} In their Table 2, ADH implement a placebo test. They estimate: a 2SLS regression of $\Delta Y_{g,1980}$ on the average of $\Delta D_{g,2000}$ and $\Delta D_{g,2010}$, using the average of $\Delta Z_{g,2000}$ and $\Delta Z_{g,2010}$ as the instrument; a 2SLS regression of $\Delta Y_{g,1990}$ on the same treatment, using the same instrument; a stacked 2SLS regression of $\Delta Y_{g,1980}$ and $\Delta Y_{g,1990}$ on the same treatment, using the same instrument. Those analyses yield a valid placebo test of Point 1 of Assumption \ref{as_commontrend_y}, if $D_{g,t}=0$ for every $t\leq 1990$. Unfortunately, as explained by ADH, trade data with China is unavailable in 1970 and 1980, so we cannot compute $D_{g,1970}$ and $D_{g,1980}$. On the other hand, $D_{g,1990}$ can be computed, and we find that it is on average equal to $0.246$: even in 1990, US CZs were on average exposed to 246 USD of imports from China per worker. The average of $D_{g,2000}$ is equal to $1.422$, which is of course larger, but maybe not by a sufficiently large order of magnitude to consider that US CZs were treated in 2000 and fully untreated in 1990. Following that logic, $\Delta Y_{g,1990}$ may not be used to test Assumptions \ref{as_commontrend_y} and \ref{as_GMM_commontrend_y}. Figure 1 in ADH shows that the import penetration ratio from China increased by 111\% from 1987 to 1990, namely a 28.2\% yearly growth rate. Extrapolating that growth rate from 1980 to 1990 would yield an average value of $D_{g,1980}$ equal to $0.021$. At the other extreme, assuming that imports from China did not grow from 1980 to 1987 would yield an average value of $D_{g,1980}$ equal to $0.117$. In the first scenario, one may argue that $\Delta Y_{g,1980}\approx \Delta Y_{g,1980}(0)$ is a reasonable approximation, while this approximation might be less reasonable in the second scenario. Accordingly, we report placebos using $\Delta Y_{g,1980}$ below, emphasizing that the absence of trade data with China in 1980 and 1970 complicates the interpretation of those tests.

\paragraph{Testing Assumptions \ref{as_GMM_commontrend_y} and \ref{as_GMM_commontrend_y_X}.} In Panel A of Table \ref{table:ADH_Placebo}, we test Assumption \ref{as_GMM_commontrend_y} by regressing $\Delta Y_{g,1980}$ on $Z_{g,1990}$. We find that CZs with a larger value of $Z_{g,1990}$ experienced a larger employment growth from 1970 to 1980, thus suggesting a positive 1970-to-1980 pre-trend, similar to that in Table 2 Column (4) of ADH. In Panel B (resp. C), we regress $\Delta Y_{g,1980}$ on $Z_{g,2000}$ (resp. $Z_{g,2010}$) and find similar results, though the magnitude of the pre-trend is smaller. Panels D to F replicate Panels A to C, adding the same control variables as in Column (6) of Table 3 of ADH, the authors' preferred specification. Those controls include census division dummies, and six ``baseline'' CZ characteristics measured in 1990. Those characteristics are CZs' percentage employment in manufacturing (manufacturing employment divided by total employment), percentage college-educated population, percentage foreign-born population, female employment rate, percentage employment in routine occupations, and average offshorability index of occupations. Those controls seem to ``kill'' the positive 1970-1980 pre-trend. Panels G to I replicate Panels D to F, keeping CZs' percentage employment in manufacturing as the only control variable and without the census division dummies. Pre-trends are no longer statistically significant. Conducting the same exercise with the remaining five control variables, we always find very significant pre-trends: percentage employment in manufacturing seems to be the key control variable to kill the pre-trend.

\paragraph{Implications.}
Without Assumption \ref{as_commontrend_y}, Theorem \ref{th_2sls_c} shows that $\theta^b$ identifies the sum of two terms: a bias term arising from the violation of Assumption \ref{as_commontrend_y}, plus a weighted sum of treatment effects. Thus, analyzing the weights in this second term is useful even if Assumption \ref{as_commontrend_y} fails, as it can help analyze a bias in $\theta^b$ that may come from heterogeneous treatment effects, on top of another bias that may come from differential trends. On the other hand, it is less straightforward to assess the impact of a violation of Assumption \ref{as_GMM_commontrend_y} on our IV-CRC estimator. Accordingly, as a robustness check we will recompute this estimator controlling for CZs' percentage employment in manufacturing, as Assumption \ref{as_GMM_commontrend_y_X} is not rejected with that control variable.\footnote{Percentage employment in manufacturing is closely related to ADH's main outcome variable. Controlling for the baseline outcome in a first-difference or fixed-effects model may lead to a so-called Nickel bias, but this is an other methodological discussion, orthogonal to that we are interested in, so we follow ADH's specification.} We also note that the pre-trend test we can run is very distant in time from the China shock. CZs' employment trends from 1970 to 1980 may not be representative of their counterfactual trends from 1990 to 2010, so our tests of Assumptions \ref{as_GMM_commontrend_y}
and \ref{as_GMM_commontrend_y_X} may not be very informative.
\begin{table}[H]
	\begin{center}
		\caption{\normalsize Pre-trends tests of Assumptions \ref{as_commontrend_y}, \ref{as_GMM_commontrend_y}, and \ref{as_GMM_commontrend_y_X}}
	\begin{tabular}{lcc}
	\hline
 	    & Estimate & Standard error \\
Regression of $\Delta Y_{g,1980}$ on:				& (1)    & (2)   \\
\hline
Panel A: $Z_{g,1990}$		 & 1.066  & 0.314 \\
Panel B: $Z_{g,2000}$		 & 0.309 &  0.099 \\
Panel C: $Z_{g,2010}$ 	& 0.117 & 0.034 \\
Panel D: $Z_{g,1990}$ and all controls in ADH & 0.162 & 0.388 \\			
Panel E: $Z_{g,2000}$ and all controls in ADH & -0.033 & 0.121 \\
Panel F: $Z_{g,2010}$ and all controls in ADH &  0.025 & 0.039 \\		
Panel G: $Z_{g,1990}$ and CZs' \% employment in manufacturing & -0.008 & 0.336 \\
Panel H: $Z_{g,2000}$ and CZs' \% employment in manufacturing & -0.113 & 0.111 \\
Panel I: $Z_{g,2010}$ and CZs' \% employment in manufacturing & -0.030 & 0.039 \\
			\hline			
			Observations & 722 & \\
			\hline
\end{tabular}
\label{table:ADH_Placebo}
\end{center}
\footnotesize{Notes: The table reports regressions using a US commuting-zone (CZ) level panel data set with five periods, 1970, 1980, 1990, 2000, and 2010. In all panels, the dependent variable is the change of the manufacturing employment per working-age population in CZ $g$, from 1970 to 1980. In Panel A, D, and G (resp. B, E, and H, C, F, and I), the main independent variable is the 1990 (resp. 2000, 2010) instrument. In Panels D to F, independent variables also include the same control variables as in Column (6) of Table 3 of \cite{autor2013china} measured in 1990 (see main text for the list of controls). In Panels G to I, independent variables also include CZs' \% employment in manufacturing. Standard errors clustered at the CZ level shown in parentheses. All regressions are unweighted.}
\end{table}

\subsubsection{Suggestive tests of Assumption \ref{as_GMM_ignorable_treatment} are conclusive}

To suggestively test Assumption \ref{as_GMM_ignorable_treatment}, which requires that $\alpha_g$ be mean independent of $D_{g,t}$ conditional on $(Z_{g,1990}, Z_{g,2000},Z_{g,2010})$, we regress the six CZ characteristics used by ADH as controls on $(D_{g,t}, Z_{g,1990}, Z_{g,2000},Z_{g,2010})$, for $t=1990$, $2000$ and $2010$. Those characteristics are likely to be correlated with CZs' effects of imports from China on their manufacturing employment. In particular, the percentage employment in routine occupations and the average offshorability index of occupations should be good predictors of $\alpha_g$. The results, shown in Table \ref{table:ADH_Covariates}, are rather conclusive. Of the 18 coefficients in Table \ref{table:ADH_Covariates}, only three are significant at the 5\% level.

\begin{table}[H]
	\begin{center}
		\caption{\normalsize Suggestive tests of Assumption \ref{as_GMM_ignorable_treatment}}
		\begin{tabular}{lcccccc}
		\hline
		& Manufacturing & College & Foreign & Women & Routine & Offshorability \\
		\hline
		$D_{g,1990}$ & -1.350 & -1.206 & -0.401 & -0.466 & -0.813 & -0.091
		 \\
		& (2.267) & (0.861) & (0.287) & (0.692) & (0.498) & (0.061) \\
		$D_{g,2000}$ & 0.060 & -0.018 & -0.139 & 0.061 & -0.111 & -0.018
		\\
		& (0.581) & (0.160) & (0.069) & (0.147) & (0.124) & (0.014) \\
		$D_{g,2010}$ & 0.603 & -0.083 & -0.098 & 0.087 & 0.040 & 0.004 \\
		& (0.296) & (0.083) & (0.031) & (0.077) & (0.063) & (0.008) \\
		\hline
		Observations & 722 & 722 & 722 & 722 & 722 & 722 \\
		\hline
		\end{tabular}
	\label{table:ADH_Covariates}
	\end{center}
\footnotesize{Notes: The table shows suggestive tests of Assumption \ref{as_GMM_ignorable_treatment}. The CZ characteristics used by \cite{autor2013china} as controls are regressed on $(D_{g,t},Z_{g,1990},Z_{g,2000},Z_{g,2010})$, for $t=1990$, $2000$, and $2010$. The table shows the coefficients of $D_{g,t}$ in those regressions, and robust standard errors. The CZ characteristics are CZs' percentage employment in manufacturing, percentage college-educated population, percentage foreign-born population, female employment rate, percentage employment in routine occupations, and average offshorability index.}
\end{table}

\subsection{Results}

\subsubsection{FD 2SLS Bartik regressions}

Columns (1) to (3) of Table \ref{table:ADH_Bartikreg} below show the results of the first-difference first-stage, reduced-form, and 2SLS Bartik regressions. In Column (3), the 2SLS coefficient is -0.564. In Column (4), the 2SLS regression is weighted by CZs' population in 1990, as in ADH, and the coefficient is -0.535. Standard errors clustered at the CZ level are shown between parentheses. All coefficients are statistically significant. The weighted 2SLS coefficient slightly differs from that in Table 2 Column (3) in ADH, because some of our variables' definitions differ, as explained above.
\begin{table}[H]
	\begin{center}
	\caption{\normalsize FD 2SLS estimates of effect of imports from China on US manufacturing employment}
		\begin{tabular}{lcccc}
		 \hline
		  & FS & RF & 2SLS & 2SLS, Weighted \\
		 & (1) & (2) & (3) & (4)  \\
          & 0.967 & -0.545 & -0.564 & -0.535 \\
			& (0.093) & (0.071) & (0.091) & (0.061) \\
			\hline
			Observations & 1,444 & 1,444 & 1,444 & 1,444 \\
			\hline
		\end{tabular}
\label{table:ADH_Bartikreg}
\end{center}
\footnotesize{Notes: Columns (1) to (3) respectively report estimates of the first-difference (FD) first-stage, reduced-form, and 2SLS Bartik regressions with period fixed effects, using a US commuting-zone (CZ) level panel data set with $T=3$ periods, 1990, 2000, and 2010. The regressions are unweighted. $\Delta Y_{g,t}$ is the change of the manufacturing employment per working-age population in CZ $g$, from 1990 to 2000 for $t=2000$, and from 2000 to 2010 for $t=2010$. $\Delta D_{g,t}$ is the change in exposure to imports from China in CZ $g$ from 1990 to 2000 for $t=2000$, and from 2000 to 2010 for $t=2010$. $\Delta Z_{g,t}$ is the first-difference Bartik instrument, whose construction is detailed in the text. Column (4) reports estimates of the first-difference 2SLS Bartik regressions, weighted by CZ's share of national population in 1990. Standard errors clustered at the CZ level shown in parentheses.} 
\end{table}

\subsubsection{Decompositions of the FD 2SLS regression}

We follow Point 1 of Theorem \ref{th_ap_2sls_c} in the Web Appendix, a generalization of Theorem \ref{th_2sls_c} to weighted FD 2SLS regressions with more than two time periods, to estimate the weights attached to the regression in Column (4) of Table \ref{table:ADH_Bartikreg}. The first column of Panel A of Table \ref{table:ADH_Bartikreg_weights} shows that $\theta^b$ estimates a weighted sum of $2166$ ($722$ CZs $\times 3$ periods) effects $\alpha_{g,t}$, where 1163 weights are positive, 1003 weights are strictly negative, and negative weights sum to $-0.734$. Therefore, $\theta^b$ is far from estimating a convex combination of effects. We do not have exactly one half of negative weights, because the regression uses three time periods, and this result is specific to the two-periods case. The weights are correlated with the year $t$ (correlation=0.082, p-value$<$0.001). We also test if the weights are correlated with the six CZ-level characteristics that ADH use as controls in their preferred specification, measured in 1990. We find that the weights are correlated with CZs' percentage employment in manufacturing (correlation=0.057, p-value=0.008), percentage foreign-born population (correlation=0.075, p-value$<$0.001), percentage employment in routine occupations (correlation=0.043, p-value=0.044), average offshorability index of occupations (correlation=0.084, p-value$<$0.001) and not significantly correlated with the other characteristics. The second column of Panel A of Table \ref{table:ADH_Bartikreg_weights} shows that even if one assumes constant effects over time, $\theta^b$ still estimates a weighted sum of $722$ location-specific effects $\alpha_{g}$, where 479 weights are strictly negative and negative weights sum to $-0.314$. As our regression is not numerically identical to the regressions in ADH, we also estimate the weights attached to the regressions in their Table 2, Column (1) and (2), under the assumption that $\alpha_{g,t}=\alpha_{g}$. The regressions in their Table 2 Column (1) and (2) only use two periods of data, thus allowing us to bypass the fact that their shares are time-varying (see \eqref{eq:deltad_def_ADH}): with only one first-difference, their shares are time-invariant, as in our decompositions (extending our decompositions to allow for time-varying shares would not be difficult). Assuming that $\alpha_{g,t}=\alpha_{g}$ allows us to bypass the fact that their first-differenced treatment is hard to reconcile with a treatment in levels (see \eqref{eq:deltad_def_ADH}): when assuming that $\alpha_{g,t}=\alpha_{g}$, our decomposition of $\theta^b$ no longer depends on $D_{g,t}$. In Panel B of Table \ref{table:ADH_Bartikreg_weights}, we find similar results as in the second column of Panel A.
\begin{table}[H]
	\begin{center}
		\caption{\normalsize Weights attached to FD 2SLS regressions}
		\begin{tabular}{lcc}

			\hline
Panel A: Regression in Column (4) of Table \ref{table:ADH_Bartikreg} & & \\
			Assumption on treatment effects & None  & $\alpha_{g,t}=\alpha_{g}$ \\
			Number of strictly negative weights & 1003  & 479 \\
			Number of positive weights  & 1163  & 243   \\
			Sum of negative weights  & -0.734  & -0.314  \\ \hline
Panel B: Regressions in ADH Table 2, assuming $\alpha_{g,t}=\alpha_{g}$ & & \\
Column \# in ADH Table 2 & (1)  & (2) \\
				Number of strictly negative weights & 454  & 429 \\
				Number of positive weights  & 268  &  293  \\
				Sum of negative weights  & -0.315  &  -0.339 \\ \hline
\end{tabular}
		\label{table:ADH_Bartikreg_weights}
\end{center}
\footnotesize{Notes: Panel A reports summary statistics on the weights attached to the first-difference (FD) 2SLS regression in Column (4) of Table \ref{table:ADH_Bartikreg}. In the first column, no assumption is made on the treatment effects. In the second column, we assume that treatment effects do not vary over time ($\alpha_{g,t}=\alpha_{g}$). Panel B reports summary statistics on the weights attached to the FD 2SLS regressions in ADH Table 2, assuming $\alpha_{g,t}=\alpha_{g}$.}
\end{table}

\subsubsection{Alternative IV-CRC estimator}

\paragraph{Main results.} In Table \ref{table:ADH_IV-CRC}, we report IV-CRC estimates of the effects of imports from China on CZs' employment, following Theorem \ref{th_chamberlain}. Our baseline specification assumes that  \begin{equation}\label{eq:cefmodel2}
E(D_{g,t}|Z_{g,1990},Z_{g,2000}, Z_{g,2010})=\delta_{0,t}+\delta_{t,t}Z_{g,t},
\end{equation}
and reports $$\widehat{\alpha}_{ate,w}\equiv \frac{1}{G}\sum_{g=1}^{722}pop_g\sum_t (\widehat{\alpha}_g+\widehat{\lambda}_t),$$
where $pop_g$ denotes CZs' populations in 1990. $\widehat{\alpha}_{ate,w}$ weights the CZ-specific effects $\sum_t (\widehat{\alpha}_g+\widehat{\lambda}_t)$ by CZs' population, consistent with the weighted FD 2SLS regression in Column (4) of Table \ref{table:ADH_Bartikreg}. Our baseline estimate is positive, small, and insignificantly different from 0. It is significantly different from the coefficient in Column (4) of Table \ref{table:ADH_Bartikreg} (t-stat=-2.015 clustering at the CZ level, t-stat=-1.712 clustering at the state level like ADH). Importantly, our IV-CRC estimator is also significantly different from the original FD 2SLS estimate in Table 2 Column (3) of ADH (t-stat=-2.675 clustering at the CZ level, t-stat=-2.291 clustering at the state level like ADH). The standard error of our IV-CRC estimate is about 5 times larger than that of the 2SLS estimate: allowing for some treatment-effect heterogeneity comes with a cost in terms of precision. Still, the confidence interval of our IV-CRC estimator does not contain the FD 2SLS estimate in Table 2 Column (3) of ADH, or that in their Table 3 Column (6).
\begin{table}[H]
	\begin{center}
	\caption{\normalsize IV-CRC estimates of the effect of imports from China on US manufacturing employment}
		\begin{tabular}{lcc}
		 \hline
		  & Estimate & Standard error \\
           & (1)  & (2) \\
Baseline estimate          & 0.138  &  0.312 \\
Estimate controlling for CZs' percentage employment in manufacturing & -0.224  & 0.319 \\
First-stage model where treatment regressed on instrument at all dates & 0.492  & 0.685 \\
Estimate assuming constant effects over time & -0.501 & 0.225 \\
			\hline
			Observations & 722 & \\
			\hline
		\end{tabular}
\label{table:ADH_IV-CRC}
\end{center}
\footnotesize{Notes: Columns (1) and (2) report IV-CRC estimates of the effect of imports from China on US manufacturing employment, computed using a US commuting-zone (CZ) level panel data set with $T=3$ periods, 1990, 2000, and 2010. $Y_{g,t}$ is the manufacturing employment per working-age population in CZ $g$ in year $t$. $D_{g,t}$ is the exposure to imports from China in CZ $g$ in year $t$. $Z_{g,t}$ is the instrument, whose construction is detailed in the text. Column (1) reports IV-CRC estimates computed following Theorem \ref{th_chamberlain}. Column (2) reports bootstrapped standard errors.}
\end{table}

\paragraph{Robustness checks.} In view of the positive 1970-to-1980 pre-trend shown in Table \ref{table:ADH_Placebo}, which disappears once CZs' 1990 percentage employment in manufacturing is controlled for, we recompute our IV-CRC estimator controlling for that variable. The second line of Table \ref{table:ADH_IV-CRC} shows that with this control, the IV-CRC estimate becomes negative, but is still fairly small and insignificant. A cross-validation exercise, where we compare the out-of-sample fit of the model in \eqref{eq:cefmodel2} and of polynomials of order 1 to 3 in $(Z_{g,1990},Z_{g,2000},Z_{g,2010})$, shows that the polynomial of order 1 with all lags and leads of the instrument has the best out-of-sample fit, closely followed by \eqref{eq:cefmodel2}, and the two models are much better than all the other models. Accordingly, we recompute our IV-CRC estimate, using a polynomial of order 1 in $(Z_{g,1990},Z_{g,2000},Z_{g,2010})$ as the first-stage model. The resulting estimate is positive, insignificant, and much more noisy than our baseline estimate. Finally, we compute an IV-CRC estimate assuming constant effects over time. Interestingly, this estimate is large, negative, significant, and very close to the FD 2SLS estimate. It is also significantly different from our baseline IV-CRC estimate (t-stat=-2.330), which implies that under Assumptions \ref{as_additivsep_FS2S}-\ref{as_GMM_ignorable_treatment}, we can reject the null that the treatment effect is constant over time. This suggests that time-varing effects might bias the FD 2SLS estimate.

\paragraph{Testing for heterogeneous effects across CZs.} To test for heterogeneous effects across CZs, we regress the CZ-specific estimated effects $\sum_t (\widehat{\alpha}_g+\widehat{\lambda}_t)$ on the
six 1990-CZ-level characteristics used by ADH as controls in their preferred specification, weighting the regression by CZs' population. To obtain standard errors, we bootstrap the whole estimation procedure, clustering the bootstrap at the CZ level. Panel A of Table \ref{table:ADH_heteffects} below shows that CZ-specific effects are significantly negatively correlated with CZs' percentage employment in manufacturing and in routine occupations, but only the latter remains significant at the 5\% level after a Bonferroni adjustment accounting for the six dimensions of heterogeneity tested in Table \ref{table:ADH_heteffects}. Then, we average $\sum_t (\widehat{\alpha}_g+\widehat{\lambda}_t)$ across CZs with an employment rate in routine occupations above and below the median, weighting the median by CZs population. It turns out that CZs' employment rate in routine occupations is highly correlated with their population, so the 72 CZs with the largest employment rate in routine occupations account for 50\% of CZs' population. Panel B of Table \ref{table:ADH_heteffects} shows that in those 72 CZs, our IV-CRC estimate is very slightly negative, but still insignificant. In the remaining CZs, our IV-CRC estimate is positive and insignificant. The difference between the IV-CRC estimates in the two subgroups is highly significant (t-stat=$-3.694$).
\begin{table}[H]
	\begin{center}
		\caption{\normalsize Heterogeneous treatment effects}
		\begin{tabular}{lcc}
		\hline
		  & Estimate & Standard error \\
			& (1) & (2) \\
\textit{Panel A: Predictors of CZs' treatment effects} & & \\
	Percentage employment in manufacturing & -0.033  & 0.015  \\
 Percentage college-educated population & -0.012 & 0.012 \\
 Percentage foreign-born population & 0.011 & 0.006 \\
 Female employment rate & 0.023 & 0.015 \\
 Percentage employment in routine occupations & -0.127 & 0.049  \\
 Average offshorability index of occupations & -0.036 & 0.210 \\
 Observations & 722 & \\ \hline
\textit{Panel B: Subgroup analysis} & & \\
Above median \% employment in routine occupations & -0.059 & 0.297 \\
Observations & 72 & \\
Below median \% employment in routine occupations & 0.336 & 0.338 \\
Observations & 650 & \\
		\hline
		\end{tabular}
	\label{table:ADH_heteffects}
	\end{center}
\footnotesize{Notes: Panel A shows results from a regression of $\sum_t (\widehat{\alpha}_g+\widehat{\lambda}_t)$, the estimated average effect of imports from China on employment of commuting-zone (CZ) $g$, on six CZ characteristics measured in 1990. The six characteristics are CZs' percentage employment in manufacturing, percentage college-educated population, percentage foreign-born population, female employment rate, percentage employment in routine occupations, and average offshorability index of occupations. Column (1) shows the coefficient of each variable in the regression, Column (2) shows a standard error, computed by bootstrapping the whole estimation procedure, clustering the bootstrap at the CZ level. The regression is run in the sample of 722 CZs used by ADH. Column (1) Panel B shows IV-CRC estimates of the effect of imports from China on the manufacturing employment share, separately for CZs above and below the median of percentage employment in routine occupations, where the median is weighted by CZs population. Panel B Column (2) shows the bootstrapped standard error of effects in Column (1).}
\end{table}

\section{Recommendations for practitioners}

In Bartik designs, we recommend that practitioners start their analysis by testing the random-shocks assumption. To do so, they can regress the shocks $Z_{s,t}$ on sectors' average share across locations $Q_{s,.}$, and/or sectoral characteristics, controlling for period fixed effects.

\medskip
When $Q_{s,.}$ and/or sectoral characteristics do not significantly predict the shocks, this is evidence that shocks are as-good-as randomly assigned. Then, practitioners may either use a slightly modified FD 2SLS Bartik estimator, where shocks are standardized by their period-specific standard deviation when constructing the instrument, or a pooled-cross-section 2SLS Bartik regression of $Y_{g,t}$ on $D_{g,t}$ using $Z_{g,t}$ as the instrument, with period fixed effects but no location fixed effects. With randomly-assigned shocks, both regressions estimate a convex combination of effects, even if effects vary over time and across locations.

\medskip
On the other hand, when $Q_{s,.}$ and/or sectoral characteristics significantly predict the shocks, this is evidence that shocks are not as-good-as randomly assigned. Then, practitioners can start by estimating the weights in the decomposition of the FD 2SLS coefficient we give in our Theorem \ref{th_2sls_c}. If most or all weights are positive, this coefficient is robust to heterogeneous treatment effects under a fairly minimal parallel trends assumption, so using that estimator may be a reasonable choice. If many weights are negative, and if weights are correlated with characteristics likely to be correlated with treatment effects, the FD 2SLS coefficient may be biased. In such instances, practitioners may consider using our IV-CRC estimator instead. If the data contains a period $t_0$ such that all locations are untreated at $t_0$ and $t_0-1$, we recommend that practitioners test the exogeneity condition underlying our IV-CRC estimator, by  regressing $\Delta Y_{g,t_0}$ on $Z_{g,t'}$ for $t' \ne t_0$. Our estimator also requires that locations' treatment effects be independent of $D_{g,t}$, conditional on $\bm{Z}_{g}$. We recommend that practitioners also suggestively test that assumption, by regressing covariates likely to be correlated with locations' treatment effects on $(D_{g,t},\bm{Z}_{g})$.

\medskip
When the instrument does not have a shift-share structure, our recommendations are similar, except that one should start by testing whether the instrument $Z_{g,t}$ is as-good-as randomly assigned. Depending on the results of that test, the same decision tree unfolds.

\newpage

\bibliography{biblio}

\newpage

\section{Proofs}

\subsection{Proof of Lemma \ref{lem:causal_fd_levels}}
\eqref{eq:causal_model_levels} implies that for any $(d,d')$ and $t\geq 2$,
\begin{align}\label{eq:steps_causal_level_fd_1}
			Y_{g,t}(d)-Y_{g,t-1}(d')=\Delta Y_{g,t}(0)+\Delta \alpha_{g,t} d'+\alpha_{g,t}(d-d').
		\end{align}
Then, if $d=d'$, $Y_{g,t}(d)-Y_{g,t-1}(d)$ is an outcome evolution without any treatment change, so
\begin{align}\label{eq:steps_causal_level_fd_2}
			Y_{g,t}(d)-Y_{g,t-1}(d)=\Delta Y_{g,t} (\tilde{0}).
		\end{align}
Combining \eqref{eq:steps_causal_level_fd_1} at $d=d'$ and \eqref{eq:steps_causal_level_fd_2},
\begin{align}\label{eq:steps_causal_level_fd_3}
			\Delta Y_{g,t}(0)+\Delta \alpha_{g,t} d=\Delta Y_{g,t} (\tilde{0}).
		\end{align}
As the right-hand-side of \eqref{eq:steps_causal_level_fd_3} does not depend on $d$, one must have that for all $t\geq 2$, $\Delta \alpha_{g,t}=0$: $\alpha_{g,t}=\alpha_{g,t-1}$. Therefore there exists $\alpha_g$ such that $\alpha_g=\alpha_{g,t}$ for all $t$. Then, it follows from \eqref{eq:steps_causal_level_fd_3} that $\Delta Y_{g,t}(0)=\Delta Y_{g,t} (\tilde{0})$, and it finally follows from \eqref{eq:causal_model_fd} and \eqref{eq:steps_causal_level_fd_1} that $\alpha^{fd}_{g,t}=\alpha_{g}$.

\subsection{Proof of Theorem \ref{th_2sls_c}}
\begin{align}\label{eq:thm3_2}
	& E \left( \sum_{g=1}^G \Delta Y_{g} \left( \Delta Z_{g}-E\left({\Delta Z}_{.}\right)\right) \right) \nonumber \\
	=& E \left( \sum_{g=1}^G \left( \Delta Y_{g}(0)+\alpha_{g,2}D_{g,2}-\alpha_{g,1}D_{g,1}\right)  \left( \Delta Z_{g}-E\left({\Delta Z}_{.}\right)\right) \right) \nonumber \\
	=& \sum_{g=1}^G E \left(  \Delta Y_{g}(0)  \left( \Delta Z_{g}-E\left({\Delta Z}_{.}\right)\right) \right)+ E \left(\sum_{g=1}^G \sum_{t=1}^2 (1\{t=2\}-1\{t=1\})D_{g,t}\left( \Delta Z_{g}-E\left({\Delta Z}_{.}\right)\right) \alpha_{g,t}\right).
\end{align}	
The first equality follows from Assumption \ref{as_linear_ss}.
Similarly,
\begin{align}\label{eq:thm3_3}
	& E \left( \sum_{g=1}^G \Delta D_{g} \left( \Delta Z_{g}-E\left({\Delta Z}_{.}\right)\right) \right) \nonumber \\
	=& E \left(\sum_{g=1}^G \sum_{t=1}^2 (1\{t=2\}-1\{t=1\})D_{g,t}\left( \Delta Z_{g}-E\left({\Delta Z}_{.}\right)\right)\right).
\end{align}	
Then, plugging \eqref{eq:thm3_2} and \eqref{eq:thm3_3} into \eqref{eq:thm3_1} yields Point 1 of the theorem. Point 2 directly follows from Point 1.
\textbf{QED.}

\subsection{Theorem \ref{th_2sls_c_fs}}
\begin{align}\label{eq:thmFS_1}
	& E \left( \sum_{g=1}^G \Delta Y_{g} \left( \Delta Z_{g}- E \left( {\Delta Z}_{.} \right) \right) \right) \nonumber \\
	=& E \left( \sum_{g=1}^G \left( \Delta Y_g (D_{g} (\bm{0}))+\alpha_{g,2} \sum_{s=1}^S Q_{s,g}  \beta_{s,g,2} Z_{s,2}- \alpha_{g,1}\sum_{s=1}^S Q_{s,g}  \beta_{s,g,1} Z_{s,1} \right)  \left( \Delta Z_{g}- E \left( {\Delta Z}_{.} \right) \right) \right) \nonumber \\
	=& \sum_{g=1}^G E \left( \Delta Y_g (D_{g} (\bm{0})) \left( \Delta Z_{g}- E \left( {\Delta Z}_{g} \right) \right)  \right) \nonumber \\
	+& E\left( \sum_{g=1}^G \sum_{t=1}^2 (1\{t=2\}-1\{t=1\})
	 \sum_{s=1}^S Q_{s,g} \beta_{s,g,t} Z_{s,t}   \left( \Delta Z_{g}- E \left( {\Delta Z}_{.} \right) \right) \alpha_{g,t} \right) \nonumber \\
	=&  E\left( \sum_{g=1}^G \sum_{t=1}^2 (1\{t=2\}-1\{t=1\})
	\sum_{s=1}^S Q_{s,g} \beta_{s,g,t} Z_{s,t}   \left( \Delta Z_{g}- E \left( {\Delta Z}_{.} \right) \right) \alpha_{g,t} \right).
\end{align}
The first equality follows from \eqref{eq:obs_outcome3}. The second equality follows from Point 3 of Assumption \ref{as_commontrend_rfy}. The third equality follows from Point 1 of Assumption \ref{as_commontrend_rfy}. Similarly, one can show that
\begin{align}\label{eq:thmFS_2}
	& E \left( \sum_{g=1}^G \Delta D_{g} \left( \Delta Z_{g}- E\left( {\Delta Z}_{.} \right) \right) \right) \nonumber \\
	=& E \left( \sum_{g=1}^G \sum_{t=1}^{2} (1\{t=2\}-1\{t=1\})  \sum_{s=1}^S Q_{s,g} \beta_{s,g,t} Z_{s,t}  \left( \Delta Z_{g}-E \left( {\Delta Z}_{.} \right) \right)  \right).
\end{align}
Then, plugging \eqref{eq:thmFS_1} and \eqref{eq:thmFS_2} into \eqref{eq:thm3_1} yields Point 1 of the theorem. Points 2 and 3 directly follows from Point 1.

\subsection{Theorem \ref{th_2sls_c_randomshocks}}
First, as shares sum to one, it follows from Assumption \ref{as_randomly_assigned_shocks} that $E(\Delta Z_g|\mathcal{F})=\Delta m$ for all $g$, so
\begin{align}\label{eq:resultAKM_1}
E(\Delta Z_.)=\Delta m.
\end{align}
Then,
\begin{align}\label{eq:resultAKM_2}
&E(\Delta Y_g(\Delta Z_g-\Delta m))\nonumber\\
=&E\left((\Delta Y_{g} (0) + \alpha_{g,2} D_{g,2}-\alpha_{g,1} D_{g,1})(\Delta Z_g-\Delta m)\right)\nonumber\\
=&E\left(\left(\Delta Y_{g} (0) + \alpha_{g,2} D_{g,2}(\bm{0})-\alpha_{g,1} D_{g,1}(\bm{0})\right)(E(\Delta Z_g|\mathcal{F})-\Delta m)\right)\nonumber\\
+&E\left(\left(\sum_{s=1}^SQ_{s,g}\beta_{s,g,2}Z_{s,2}(\Delta Z_g-\Delta m)\right)\alpha_{g,2}\right)-E\left(\left(\sum_{s=1}^SQ_{s,g}\beta_{s,g,1}Z_{s,1}(\Delta Z_g-\Delta m)\right)\alpha_{g,1}\right)\nonumber\\
=&E\left(\left(\sum_{s,s'}Q_{s,g}Q_{s',g}\beta_{s,g,2}E\left(Z_{s,2}(Z_{s',2}-m_2)|\mathcal{F}\right)-\sum_{s,s'}Q_{s,g}Q_{s',g}\beta_{s,g,2}E\left(Z_{s,2}(Z_{s',1}-m_1)|\mathcal{F}\right)\right)\alpha_{g,2}\right)\nonumber\\
+&E\left(\left(\sum_{s,s'}Q_{s,g}Q_{s',g}\beta_{s,g,1}E\left(Z_{s,1}(Z_{s',1}-m_1)|\mathcal{F}\right)-\sum_{s,s'}Q_{s,g}Q_{s',g}\beta_{s,g,1}E\left(Z_{s,1}(Z_{s',2}-m_2)|\mathcal{F}\right)\right)\alpha_{g,1}\right)\nonumber\\
=&E\left(\sum_{s=1}^SQ^2_{s,g}\beta_{s,g,2}\left(V\left(Z_{s,2}\right)-\cov\left(Z_{s,1},Z_{s,2}\right)\right)\alpha_{g,2}\right)\nonumber\\
+&E\left(\sum_{s=1}^SQ^2_{s,g}\beta_{s,g,1}\left(V\left(Z_{s,1}\right)-\cov\left(Z_{s,1},Z_{s,2}\right)\right)\alpha_{g,1}\right).
\end{align}
The first equality follows from \eqref{eq:obs_outcome}. The second equality follows from \eqref{eq:obs_treat} and the law of iterated expectations. The third equality follows from the fact  that $E(\Delta Z_g|\mathcal{F})=\Delta m$, by Assumption \ref{as_randomly_assigned_shocks} and as shares sum to one, and from the law of iterated expectations. The fourth equality follows from the fact that by Assumption \ref{as_randomly_assigned_shocks}, for all $s\ne s'$ and $(t,t')\in \{1,2\}^2$, $E\left(Z_{s,t}(Z_{s',t'}-m_{t'})|\mathcal{F}\right)=E\left(Z_{s,t}|\mathcal{F}\right)E\left(Z_{s',t'}-m_{t'}|\mathcal{F}\right)=0$. Moreover, as for all $s$ and $(t,t')\in \{1,2\}^2$, $E\left(Z_{s,t}Z_{s,t'}|\mathcal{F}\right)=E\left(Z_{s,t}Z_{s,t'}\right)$ and $E\left(Z_{s,t}|\mathcal{F}\right)=m_t$, $E\left(Z_{s,t}(Z_{s',t'}-m_{t'})|\mathcal{F}\right)=\cov\left(Z_{s,t},Z_{s,t'}\right)$.

\medskip
Similarly, one can show that
\begin{align}\label{eq:resultAKM_3}
&E(\Delta D_g(\Delta Z_g-\Delta m))\nonumber\\
=&E\left(\sum_{s=1}^SQ^2_{s,g}\beta_{s,g,2}\left(V\left(Z_{s,2}\right)-\cov\left(Z_{s,1},Z_{s,2}\right)\right)\right)\nonumber\\
+&E\left(\sum_{s=1}^SQ^2_{s,g}\beta_{s,g,1}\left(V\left(Z_{s,1}\right)-\cov\left(Z_{s,1},Z_{s,2}\right)\right)\right).
\end{align}
The result follows plugging \eqref{eq:resultAKM_1}, \eqref{eq:resultAKM_2}, and \eqref{eq:resultAKM_3} into \eqref{eq:thm3_1}.

\subsection{Theorem \ref{th_chamberlain}}
For all $g$ and $t$,
	\begin{align*}
		E(Y_{g,t}|\bm{Z}_{g})=&E(Y_{g,t}(0)| \bm{Z}_{g})+E(\alpha_{g} D_{g,t}| \bm{Z}_{g})+1\{t\geq 2\}\lambda_t E(D_{g,t}| \bm{Z}_{g})\nonumber\\
=&E(Y_{g,1}(0)| \bm{Z}_{g})+1\{t\geq 2\}\mu_{1:t}+(E(\alpha_{g}| \bm{Z}_{g})+1\{t\geq 2\}\lambda_t)E( D_{g,t}| \bm{Z}_{g})\nonumber\\
=&1\{t\geq 2\}(\mu_{1:t}+\lambda_t\tilde{D}_{g,t})+E(Y_{g,1}(0)| \bm{Z}_{g})+E(\alpha_{g}| \bm{Z}_{g}) \tilde{D}_{g,t}\nonumber
	\end{align*}
The first equality follows from Assumptions \ref{as_linear_ss} and \ref{as_additivsep_FS2S}, the second equality follows from  Assumption \ref{as_GMM_commontrend_y} and from the law of iterated expectations and Assumption \ref{as_GMM_ignorable_treatment}.
The previous display and the law of iterated expectations imply that
\begin{align*}
		E(Y_{g,t}|\bm{\tilde{D}}_{g})=1\{t\geq 2\}(\mu_{1:t}+\lambda_t\tilde{D}_{g,t})+E(Y_{g,1}(0)| \bm{\tilde{D}}_{g})+E(\alpha_{g}| \bm{\tilde{D}}_{g}) \tilde{D}_{g,t}.
	\end{align*}
Let $\bm{\gamma}_g=(E(Y_{g,1}(0)|\bm{\tilde{D}}_{g}),E(\alpha_g|\bm{\tilde{D}}_{g}))'$. It follows from the previous display that
	\begin{equation}\label{eq:thm5}
		E( \bm{ Y}_{g}| \bm{\tilde{D}}_{g})=\mathcal{P}_g\bm{\theta}+\mathcal{X}_{g}\bm{\gamma}_g.
	\end{equation}
As $M(\mathcal{X}_{g})\mathcal{X}_{g}=0$, left-multiplying \eqref{eq:thm5} by $\mathcal{P}_g'M(\mathcal{X}_{g})$,
\begin{equation*}
E(\mathcal{P}_g'M(\mathcal{X}_{g})\bm{Y}_{g}|\bm{\tilde{D}}_{g})=\mathcal{P}_g'M(\mathcal{X}_{g})\mathcal{P}_g\bm{\theta}.
\end{equation*}
Therefore, by the law of iterated expectation and averaging across locations:
\begin{align*}
	E\left( \frac{1}{G} \sum_{g=1}^G \mathcal{P}_g'M(\mathcal{X}_{g})\bm{Y}_{g} \right)=E \left( \frac{1}{G} \sum_{g=1}^G \mathcal{P}_g'M(\mathcal{X}_{g})\mathcal{P}_g \right) \bm{\theta}.
\end{align*}
\eqref{eq:thetaY} follows from the previous display and the fact $E\left(\frac{1}{G}\sum_{g=1}^G \mathcal{P}_g'M(\mathcal{X}_{g})\mathcal{P}_g\right)$ is invertible.

\medskip
Then, we left-multiply \eqref{eq:thm5} by $\mathcal{X}_{g}'$, and it follows that
\begin{equation*}
	E(\mathcal{X}_{g}' \bm{Y}_{g}| \bm{\tilde{D}}_{g})=\mathcal{X}_{g}' \mathcal{P}_g\bm{\theta}+\mathcal{X}_{g}'\mathcal{X}_{g}\gamma_g.
\end{equation*}
\eqref{eq:alphag} follows from: rearranging; the fact $\mathcal{X}_{g}'\mathcal{X}_{g}$ is invertible with probability one; the law of iterated expectations; and averaging across locations.

\appendix

\newpage

\LARGE

\textbf{Web Appendix: not for publication}

\normalsize

\section{Economic interpretation of our econometric assumptions in ADH, under a gravity-based decomposition of trade flows.}\label{sec_theorymetrics}

\paragraph{A gravity-based decomposition of trade flows.}
Assume that
\begin{align}\label{eq:decompo_supply_demand}
M^{US}_{s,t}=S_{s,t}+\text{Dem}^{US}_{s,t}\nonumber\\
M^{OC}_{s,t}=S_{s,t}+\text{Dem}^{OC}_{s,t},
\end{align}
where $S_{s,t}$ denotes China's
export-supply capabilities in $s$ at $t$, and $Dem^{US}_{s,t}$ and $Dem^{OC}_{s,t}$ respectively denote demand's contribution to China's exports to the US and to other countries. Under the assumptions outlined in \cite{arkolakis2012new}, if exports are in logs,\footnote{ADH use exports instead of the log of exports in their empirical analysis, while their Equation (2) is a log-log equation. To keep our empirical specification as close as possible to theirs, we too use exports instead of the log of exports in our empirical analysis, despite the fact our econometric assumptions are easier to interpret in a gravity-based framework with the log of exports.}
\begin{align}\label{eq:supply_demand_terms}
S_{s,t}=&(1-\sigma_{s,t})(\ln(w^C_{s,t}+\ln(\tau^C_{s,t}))\nonumber\\
\text{Dem}^{US}_{s,t}=&\ln(\text{Exp}^{US}_{s,t})-(1-\sigma_{s,t})\ln(P^{US}_{s,t})\nonumber\\
\text{Dem}^{OC}_{s,t}=&\ln(\text{Exp}^{OC}_{s,t})-(1-\sigma_{s,t})\ln(P^{OC}_{s,t}),
\end{align}
where $\sigma_{s,t}$ is the elasticity of substitution in sector $s$ at $t$, $w^C_{s,t}$ is the wage in China in $s$ at $t$, $\tau^C_{s,t}$ is China's variable cost of trade in $s$ at $t$,\footnote{This variable cost may not be the same when China exports to the US and to other countries. To account for that, one could allow China's export-supply capabilities to depend on the destination, without changing the economic interpretation of our econometric assumptions.} $\text{Exp}^{US}_{s,t}$ and $\text{Exp}^{OC}_{s,t}$ are the expenditures in $s$ at $t$ in the US and in other countries, and $P^{US}_{s,t}$ and $P^{OC}_{s,t}$ are the price index in $s$ at $t$ in the US and in other countries \citep[see][]{dixit1977monopolistic}.

\paragraph{Sufficient conditions for Assumption \ref{as_commontrend_y} under \eqref{eq:decompo_supply_demand}.}
With time-invariant US sectoral employments $E_{s}$ and shares $Q_{s,g}$, under \eqref{eq:decompo_supply_demand} one has
\begin{equation*}\label{eq:instrum_obs_Autor2}
	\Delta Z_{g,t}=\sum_{s=1}^S Q_{s,g} \frac{\Delta S_{s,t}+\Delta\text{Dem}^{OC}_{s,t}}{E_{s}}.
\end{equation*}
Assume that
\begin{equation}\label{eq:indep_supplydemandshocks}
(\Delta S_{s,t},\Delta\text{Dem}^{OC}_{s,t})_{s\in \{1,...,S\},t\in \{2,...,T\}}\indep ((Q_{s,g})_{s\in \{1,...,S\}},\Delta Y_{g,t}(0))_{g\in \{1,...,G\},t\in \{2,...,T\}},
\end{equation}
meaning that the first-differences of China's export supply capabilities and of other-countries demand's shocks are independent of US CZs sectoral shares and potential employment evolutions without imports from China. Then, if
$(E_{s})_{s\in \{1,...,S\}}$ is non-stochastic and for all $(s,g,t)$ $\cov(Q_{s,g},\Delta Y_{g,t}(0))=0,$
\begin{align*}
cov(\Delta Z_{g,t},\Delta Y_{g,t}(0))=&\sum_{s=1}^S cov\left(Q_{s,g} \frac{\Delta S_{s,t}+\Delta\text{Dem}^{OC}_{s,t}}{E_{s}},\Delta Y_{g,t}(0)\right)\\
=&\sum_{s=1}^S E\left(\frac{\Delta S_{s,t}+\Delta\text{Dem}^{OC}_{s,t}}{E_{s}}\right)cov\left(Q_{s,g} ,\Delta Y_{g,t}(0)\right)\\
=&0,
\end{align*}
thus providing an economic justification of Assumption \ref{as_commontrend_y} in the spirit of the shares approach of \cite{goldsmith2020bartik}.
Similarly, one can show that if shares sum to 1, \eqref{eq:indep_supplydemandshocks} holds, $(E_{s})_{s\in \{1,...,S\}}$ is non-stochastic, and $$E\left(\frac{\Delta S_{s,t}+\Delta\text{Dem}^{OC}_{s,t}}{E_{s}}\right)=m_t,$$ then
$\cov(\Delta Z_{g,t},\Delta Y_{g,t}(0))=0$, thus providing an economic justification of Assumption \ref{as_commontrend_y} in the spirit of the shocks approach of \cite{borusyak2020quasi}. With stochastic US sectoral employments $(E_{s})_{s\in \{1,...,S\}}$, rationalizing Point 2 of Assumption \ref{as_commontrend_y} would require replacing \eqref{eq:indep_supplydemandshocks} by
\begin{equation*}
(E_{s},\Delta S_{s,t},\Delta\text{Dem}^{OC}_{s,t})_{s\in \{1,...,S\},t\in \{2,...,T\}}\indep ((Q_{s,g})_{s\in \{1,...,S\}},\Delta Y_{g,t}(0))_{g\in \{1,...,G\},t\in \{2,...,T\}},
\end{equation*}
which is much less plausible: US sectoral employments are very likely to be correlated with US CZs' counterfactual employment evolutions $\Delta Y_{g,t}(0)$. This motivates using pre-determined sectoral employments to construct the instrument, as we do when we revisit ADH.

\paragraph{Sufficient condition for Assumption \ref{as_GMM_commontrend_y} under \eqref{eq:decompo_supply_demand}.}
With time-invariant US sectoral employments and shares, under \eqref{eq:decompo_supply_demand} one has
\begin{equation*}\label{eq:instrum_obs_Autor3}
	Z_{g,t}=\sum_{s=1}^S Q_{s,g} \frac{S_{s,t}+\text{Dem}^{OC}_{s,t}}{E_{s}}.
\end{equation*}
Assume that
\begin{equation}\label{eq:indep_supplydemandshocks2}
(S_{s,t},\text{Dem}^{OC}_{s,t})_{s\in \{1,...,S\},t\in \{1,...,T\}}\indep ((Q_{s,g})_{s\in \{1,...,S\}},\Delta Y_{g,t}(0))_{g\in \{1,...,G\},t\in \{2,...,T\}},
\end{equation}
meaning that China's export supply capabilities and demand's contribution to China's exports to other countries are independent of US CZs sectoral shares and potential employment evolutions without imports from China. Then, if
$(E_{s})_{s\in \{1,...,S\}}$ is non-stochastic, and for all $(g,t)$ $$E\left(\Delta Y_{g,t}(0)|(Q_{s,g})_{s\in \{1,...,S\}}\right)=\mu_t,$$ Assumption \ref{as_GMM_commontrend_y} holds.

\paragraph{Non-causal first-stage.}
Under \eqref{eq:decompo_supply_demand}, the shocks $Z_{s,t}$ do not have a direct causal effect on $D_{g,t}$. Rather, $M^{OC}_{s,t}$ and $M^{US}_{s,t}$ are co-determined by China's
export-supply capabilities $S_{s,t}$, thus leading to a statistical but non-causal first-stage between $Z_{g,t}$ and $D_{g,t}$. Our decompositions of $\theta^b$ in Theorem \ref{th_2sls_c} do not rely on any first-stage assumption, so they hold irrespective of whether the first-stage is causal or not. Similarly, our IV-CRC estimator does not rely on a causal first-stage model. Theorems \ref{th_2sls_c_fs} and \ref{th_2sls_c_randomshocks} on the other hand do rely on the causal first-stage model in Assumption \ref{as_linear_fs}. This is not an issue, as we do not use those theorems when we revisit ADH.

\section{Second empirical application: canonical Bartik design}\label{sec_applicanonical}


In this section, we revisit the canonical application in \citet{bartik1991benefits}, where the Bartik instrument is used to estimate the inverse elasticity of labor supply.

\subsection{Data}

Our data construction closely follows \citet{goldsmith2020bartik}. We construct a decennial continental US commuting-zone (CZ) level panel data set, from 1990 to 2010, with CZ wages and employment levels.
For 1990 and 2000, we use the 5\% IPUMS sample of the U.S. Census. For 2010, we pool the 2009-2011 ACSs (\citealt{usa2019}). Sectors are IND1990 sectors. We follow \citet{autor2013growth} to reallocate Public Use Micro Areas level observations of Census data to the CZ level. We also follow ADH to aggregate the Census sector code ind1990 to a balanced panel of sectors for the 1990 and 2000 Censuses and the 2009-2011 ACS, with new sector code ind1990dd.\footnote{Crosswalk files are available online at https://www.ddorn.net/data.htm. The original crosswalk file for sector code only creates a balanced panel of sectors up to the 2006-2008 ACSs. We extend the crosswalk approach to one additional sector (shoe repair shops, crosswalked into miscellaneous personal services) to create a balanced panel of sectors up to the 2009-2011 ACSs.} In our final dataset, we have 3 periods, 722 CZs and 212 sectors.

\medskip
The outcome variable $\Delta Y_{g,t}$ is the change in log wages in CZ $g$ from $t-10$ to $t$, for $t\in \{2000,2010\}$. The treatment variable $\Delta D_{g,t}$ is the change in log employment in CZ $g$ from $t-10$ to $t$. We use people aged 18 and older who are employed and report usually working at least 30 hours per week in the previous year to generate employment and average wages. We define $Q_{s,g}$ as the employment share of sector $s$ in CZ $g$ in 1990, and then construct the first-difference Bartik instrument using 1990-2000 and 2000-2010 sectoral employment growth rates.\footnote{We do not use leave-one-out growth rates, because doing so would lead to inconsistent Bartik and first-difference Bartik instruments. \citet{adao2019shift} and \citet{goldsmith2020bartik} recommend using leave-one-out to construct the national growth rates, in order to avoid the finite sample bias that comes from using own-observation information. In practice, because we have 722 locations, whether one uses leave-one-out or not to estimate the national growth rates barely changes the results.}

\subsection{Results}
\subsubsection{FD 2SLS Bartik regressions}
Columns (1) to (3) of Table \ref{table:canonical_Bartikreg} below show the results of the FD first-stage, reduced-form, and 2SLS Bartik regressions. In Column (1), the first-stage coefficient is 0.824. In Column (2), the reduced form coefficient is 0.391. Finally, in Column (3), the 2SLS coefficient is 0.475. If interpreted causally, this 2SLS coefficient means that a 1\% increase in employment leads to a 0.475\% increase in wages. Robust standard errors clustered at the CZ level are shown between parentheses. All coefficients are statistically significant.
\begin{table}[H]
	\begin{center}
		\caption{\normalsize First-difference 2SLS estimates in the canonical Bartik design}
		\begin{tabular}{lccc}
			\hline
			& FS & RF & 2SLS  \\			
			& (1) & (2) & (3)   \\
			& 0.824 & 0.391 & 0.475  \\
			& (0.055) & (0.031) & (0.039)  \\
			\hline
			Observations & 1,444 & 1,444 & 1,444  \\ \hline
		\end{tabular}
		\label{table:canonical_Bartikreg}
	\end{center}
	\footnotesize{Notes: Columns (1) to (3) respectively report estimates of first-difference 2SLS regressions with period fixed effects, using a decennial US commuting-zone (CZ) level panel data set from 1990 to 2010. $\Delta Y_{g,t}$ is the change in log wages in CZ $g$ from $t-10$ to $t$, for $t\in \{2000,2010\}$. $\Delta D_{g,t}$ is the change in log employment in CZ $g$ from $t-10$ to $t$. $\Delta Z_{g,t}$ is the first-differenced Bartik instrument, whose construction is detailed in the text. Standard errors clustered at the CZ level shown in parentheses.}
\end{table}

\subsubsection{Decompositions of the FD 2SLS regressions}

We follow Theorem \ref{th_ap_2sls_c} in the Web Appendix, a straightforward generalization of Theorem \ref{th_2sls_c} to more than two periods, to estimate the weights attached to the FD 2SLS regression under Assumptions \ref{as_linear_ss} and \ref{as_ap_commontrend_y} (the latter is a generalization of Assumption \ref{as_commontrend_y} to more than two periods). Column (1) of Table \ref{table:canonical_Bartikreg_weights} shows that under those assumptions, $\theta^b$ estimates a weighted sum of $2166$ ($722$ CZs $\times 3$ periods) effects $\alpha_{g,t}$, where 1035 weights are positive, 1131 weights are strictly negative, and where negative weights sum to $-163.495$. Therefore, $\theta^b$ is extremely far from estimating a convex combination of effects. Column (2) shows that even if one further assumes constant effects over time, $\theta^b$ still estimates a weighted sum of $722$ location-specific effects $\alpha_{g}$, where 519 weights are positive, 203 weights are strictly negative, and where negative weights sum to $-0.282$.

\begin{table}[H]
	
	\begin{center}
		\caption{\normalsize Summary statistics on the weights attached to the first-difference 2SLS regressions in Table \ref{table:canonical_Bartikreg}}
			\begin{tabular}{lcc}
				\hline
				 \\
			Assumption on treatment effects & None  & $\alpha_{g,t}=\alpha_{g}$  \\
			& (1) & (2)  \\
			Number of strictly negative weights & 1035  & 203 \\
			Number of positive weights  & 1131  & 519 \\
			Sum of negative weights  & -163.495  & -0.282  \\
			\\ \hline
		\end{tabular}
		\label{table:canonical_Bartikreg_weights}
	\end{center}
	\footnotesize{Notes: The table reports summary statistics on the weights attached to the 2SLS regression in Column (3) of Table \ref{table:canonical_Bartikreg}. The weights are estimated following Theorem \ref{th_ap_2sls_c}. In Column (1), no assumption is made on the first-stage and treatment effects. Column (2) assumes that the treatment effects do not vary over time ($\alpha_{g,t}=\alpha_{g}$).}
\end{table}

\subsubsection{Alternative IV-CRC estimator}

\paragraph{Estimation procedure.} In Table \ref{table:BTK_IV-CRC}, we follow Theorem \ref{th_chamberlain} and present IV-CRC estimates. We assume that Assumption \ref{as_GMM_functional_form} holds with $K=1$ and $K=2$, based on a cross-validation exercise, where we compare the out-of-sample fit of the models with polynomials of order 1 to 5. A standard error is obtained by bootstrapping the whole estimation procedure, clustering at the CZ level.

\paragraph{Results.}
Our IV-CRC estimate is equal to $0.316$ when $K=1$ and equal to $0.407$ when $K=2$, slightly below the 2SLS coefficient in Column (3) of Table \ref{table:canonical_Bartikreg}. Estimates' standard errors are respectively 67\% and 33\% larger than that of the FD 2SLS estimator.

\begin{table}[H]
	\begin{center}
		\caption{\normalsize IV-CRC estimates of the canonical setting}
		\begin{tabular}{lc}
			\hline
			& IV-CRC  \\
			Polynomial of order 1 in instruments at all dates          &  0.316 \\
			& (0.065) \\
			Polynomial of order 2 in instruments at all dates           &  0.407 \\
			& (0.052) \\
			\hline
			Observations & 722 \\
			\hline
		\end{tabular}
		\label{table:BTK_IV-CRC}
	\end{center}
	\footnotesize{Notes: The table reports IV-CRC estimates of the effect of employment on wages, computed using a US commuting-zone (CZ) level panel data set with $T=3$ periods, 1990, 2000, and 2010. $Y_{g,t}$ is the log wages in CZ $g$ in year $t$. $D_{g,t}$ is the log employment in CZ $g$ in year $t$. $Z_{g,t}$ is the Bartik instrument, whose construction is detailed in the text. The estimates are computed following Theorem \ref{th_chamberlain}. Bootstrapped standard error are shown in parentheses. We use two first-stage models, one where the treatment is regressed on a first-order polynomial of instruments at all dates, and one where the treatment is regressed on a second-order polynomial of instruments at all dates.}
\end{table}

\section{Weighted FD 2SLS regressions with multiple periods}\label{sec_multipleperiods}

In this section, we use the same notation and definitions as in Section \ref{sec_alternativeestimator} of the paper and we extend our decompositions of FD 2SLS regressions in Theorem \ref{th_2sls_c} to weighted regressions, with multiple periods.

\paragraph{Estimator and estimand.} With several periods, the analog of the FD 2SLS regression with a constant in the paper is a first-differenced 2SLS regression with period fixed effects. Let $w_{g,t}$ be the positive weights used in the regression, which are treated as non-stochastic quantities in what follows. For every $t$, let ${\Delta Z}^w_{.,t}=\frac{\sum_{g=1}^G w_{g,t}\Delta Z_{g,t}}{\sum_{g=1}^G w_{g,t}}$ denote the weighted average of $\Delta Z_{g,t}$ at period $t$.
\begin{definition}\label{def_ap_2sls_c}
	FD 2SLS regression with multiple periods: let
	\begin{equation}\label{eq:ap_estimator}
		\hat{\theta}^{b}=\frac{\sum_{g=1}^G \sum_{t=2}^T w_{g,t}\Delta Y_{g,t} \left( \Delta Z_{g,t}-{\Delta Z}^w_{.,t}\right) }{\sum_{g=1}^G \sum_{t=2}^T w_{g,t}\Delta D_{g,t} \left( \Delta Z_{g,t}-{\Delta Z}^w_{.,t}\right) }
	\end{equation}
	\begin{equation}\label{eq:ap_thm3_1}
		\theta^{b}=\frac{\sum_{g=1}^G \sum_{t=2}^T w_{g,t}E\left(\Delta Y_{g,t} \left( \Delta Z_{g,t}- E\left( {\Delta Z}^w_{.,t} \right) \right) \right) }{ \sum_{g=1}^G \sum_{t=2}^T w_{g,t}E\left(\Delta D_{g,t} \left( \Delta Z_{g,t}-E\left( {\Delta Z}^w_{.,t}\right) \right) \right) }.
	\end{equation}
\end{definition}

\paragraph{Identifying assumptions.} The following assumption generalizes Assumption \ref{as_commontrend_y} to the case with multiple periods.

\begin{assumption}\label{as_ap_commontrend_y}
	\begin{enumerate}
		\item For all $g \in \{1,...,G\}$, $t \in \{2,...,T\}$, $\cov(\Delta Z_{g,t},\Delta Y_{g,t} (0))=0$.
		\item For all $t \in \{2,...,T\}$, $E\left({\Delta Z}_{g,t}\right)$ does not depend on $g$.
	\end{enumerate}
\end{assumption}

\paragraph{Decompositions of $\theta^b$ under Assumptions \ref{as_linear_ss} and \ref{as_ap_commontrend_y}.}

\begin{theorem}\label{th_ap_2sls_c}
	Suppose Assumptions \ref{as_linear_ss} and \ref{as_ap_commontrend_y} hold.
	\begin{enumerate}
	\item	Then,
\small
\begin{adjustwidth}{-105pt}{-90pt}	
\begin{align*}
		\theta^{b}=E \left( \sum_{g=1}^G \sum_{t=1}^T \frac{D_{g,t}\left(1\{t\geq 2\}w_{g,t}\left(\Delta Z_{g,t}-E\left( {\Delta Z}^w_{.,t} \right)\right)-1\{t\leq T-1\}w_{g,t+1}\left(\Delta Z_{g,t+1}-E\left({\Delta Z}^w_{.,t+1} \right) \right)\right)}{ E \left( \sum_{g'=1}^G \sum_{t'=1}^T D_{g',t'}\left(1\{t'\geq 2\}w_{g',t'}\left(\Delta Z_{g',t'}-E\left( {\Delta Z}_{.,t'} \right)\right)-1\{t'\leq T-1\}w_{g',t'+1} \left(\Delta Z_{g',t'+1}-E\left({\Delta Z}_{.,t'+1} \right) \right)\right)\right)}\alpha_{g,t}\right).
	\end{align*}
\end{adjustwidth}
\normalsize
	\item If one further assumes that for all $(g,t)$, there exists $\alpha_g$ such that $\alpha_{g,t}=\alpha_{g}$, and if the weights $w_{g,t}$ are time invariant, then
		\begin{align*}
		\theta^{b}=E \left( \sum_{g=1}^G \frac{w_{g}\sum_{t=2}^T
		\Delta D_{g,t}\left( \Delta Z_{g,t}-E\left( {\Delta Z}^w_{.,t} \right) \right)  }{ E \left( \sum_{g'=1}^G w_{g'}\sum_{t=2}^T \Delta D_{g',t}\left( \Delta Z_{g',t}-E\left( {\Delta Z}^w_{.,t} \right) \right) \right)} \alpha_{g}\right).
	\end{align*}
	\end{enumerate}
\end{theorem}

\newpage
\section{Proofs of results in Web Appendix}

\subsection{Theorem \ref{th_ap_2sls_c}}
\begin{adjustwidth}{-50pt}{-50pt}
\begin{align}\label{eq:ap_thm3_2}
	& E \left( \sum_{g=1}^G \sum_{t=2}^T w_{g,t}\Delta Y_{g,t} \left( \Delta Z_{g,t}-E \left( {\Delta Z}^w_{.,t} \right) \right) \right) \nonumber \\
	=& E \left( \sum_{g=1}^G \sum_{t=2}^T w_{g,t}\left( \Delta Y_{g,t}(0)+\alpha_{g,t} D_{g,t} -\alpha_{g,t-1} D_{g,t-1}\right)  \left( \Delta Z_{g,t}-E \left( {\Delta Z}^w_{.,t} \right) \right) \right) \nonumber \\
	=& \sum_{g=1}^G \sum_{t=2}^T w_{g,t}E \left(  \Delta Y_{g,t}(0)  \left( \Delta Z_{g,t}- E\left( {\Delta Z}_{g,t}\right) \right) \right) \nonumber \\
	+& E \left( \sum_{g=1}^G \sum_{t=2}^T w_{g,t}\alpha_{g,t} D_{g,t} \left( \Delta Z_{g,t}-E\left( {\Delta Z}^w_{.,t} \right) \right)-\sum_{g=1}^G \sum_{t=2}^T w_{g,t}\alpha_{g,t-1} D_{g,t-1} \left( \Delta Z_{g,t}-E\left( {\Delta Z}^w_{.,t} \right) \right)\right) \nonumber \\
	=& E \left( \sum_{g=1}^G \sum_{t=1}^T \left( 1\{t \geq 2\} w_{g,t}\left( \Delta Z_{g,t}-E\left( {\Delta Z}^w_{.,t} \right) \right) -1\{t \leq T-1\} w_{g,t+1}\left( \Delta Z_{g,t+1}-E\left( {\Delta Z}^w_{.,t+1} \right) \right) \right)  D_{g,t} \alpha_{g,t}  \right).
\end{align}	
\end{adjustwidth}
The first equality follows from Assumption \ref{as_linear_ss}. The second equality follows from Point 2 of Assumption \ref{as_ap_commontrend_y}. The third equality follows from Point 1 of Assumption \ref{as_ap_commontrend_y}. Similarly,
\begin{adjustwidth}{-50pt}{-50pt}
\begin{align}\label{eq:ap_thm3_3}
	& E \left( \sum_{g=1}^G \sum_{t=2}^T \Delta D_{g,t} \left( \Delta Z_{g,t}-E \left( {\Delta Z}^w_{.,t} \right) \right) \right) \nonumber \\
	=& E \left( \sum_{g=1}^G \sum_{t=1}^T \left( 1\{t \geq 2\} w_{g,t}\left( \Delta Z_{g,t}-E\left( {\Delta Z}^w_{.,t} \right) \right) -1\{t \leq T-1\} w_{g,t+1}\left( \Delta Z_{g,t+1}-E\left( {\Delta Z}^w_{.,t+1} \right) \right) \right)  D_{g,t}   \right).
\end{align}	
\end{adjustwidth}
Then, plugging \eqref{eq:ap_thm3_2} and \eqref{eq:ap_thm3_3} into \eqref{eq:ap_thm3_1} yields the result. Point 2 follows from Point 1.

\end{document}